\documentclass[12pt]{article}
\usepackage{latexsym,amsmath,amsthm,amssymb,amsfonts,amsbsy,calrsfs,multibox}

\newtheorem{theorem}{Theorem}
\newtheorem{lemma}{Lemma}

\newtheorem{prop}{Proposition}
\def\a{\alpha}
\def\b{\beta}
\def\g{\gamma}
\def\c{\gamma}
\def\d{\delta}

\def\l{\lambda}
\def\L{\Lambda}

\def\D{\Delta}
\def\m{\mu}
\def\n{\nu}
\def\N{\nabla}
\def\r{\rho}
\def\o{\omega}
\def\s{\sigma}

\def\t{\tau}

\def\x{\xi}
\def\z{\zeta}
\def\e{\varepsilon}
\def\pa{\partial}
\def\6{\partial}

\def\be{\begin{equation}}
\def\ee{\end{equation}}
\def\bea{\begin{eqnarray}}
\def\eea{\end{eqnarray}}

\def\cg{{\cal G}}

\def\cl{{\cal L}}

\def\co{{\cal O}}

\def\cy{{\cal Y}}

\newcommand{\Rn}{{\mathbb{R}}}

\def\ba{\begin{array}}
\def\ba{\end{array}}

\begin{document}
\vspace{.6cm}

\begin{centering}

{\Large {\bf On The Uniqueness of Minimal Coupling\\[15pt] in Higher-Spin Gauge Theory}}

\begin{center}
{Nicolas Boulanger$^{*,}$\footnote{Work supported by a ``Progetto Italia'' fellowship. E-mail address:
\tt{nicolas.boulanger@sns.it}}, Serge Leclercq$^{\dagger,}$\footnote{
E-mail address: \tt{serge.leclercq@umh.ac.be}}
and Per Sundell$^{*,}$\footnote{Also affiliated to INFN. E-mail address: \tt{p.sundell@sns.it}} }
\end{center}

{\small{
\begin{center}
$^*$
Scuola Normale Superiore\\
Piazza dei Cavalieri 7, 56126 Pisa, Italy\\
\vspace*{.3cm}
$^\dagger$
Service de M\'ecanique et Gravitation\\
Universit\'e de Mons-Hainaut, Acad\'emie Wallonie-Bruxelles\\
6 avenue du Champ de Mars, 7000 Mons, Belgium\\
\end{center}}}

\end{centering}

\vspace*{.5cm}

\begin{abstract}
We address the uniqueness of the minimal couplings between higher-spin fields
and gravity. These couplings are cubic vertices built from gauge non-invariant
connections that induce non-abelian deformations of the gauge algebra. We show
that Fradkin-Vasiliev's cubic $2-s-s$ vertex, which contains up to $2s-2$
derivatives dressed by a cosmological constant $\L$, has a limit where: {(i)}
$\L\rightarrow 0$; {(ii)} the spin-2 Weyl tensor scales \emph{non-uniformly}
with $s$; and {(iii)} all lower-derivative couplings are scaled away. For
$s=3$ the limit yields the unique non-abelian spin $2-3-3$ vertex found
recently by two of the authors, thereby proving the \emph{uniqueness} of the
corresponding FV vertex. We extend the analysis to $s=4$ and a class of spin
$1-s-s$ vertices. The non-universality of the flat limit high-lightens not
only the problematic aspects of higher-spin interactions with $\L=0$ but also
the strongly coupled nature of the derivative expansion of the fully nonlinear
higher-spin field equations with $\L\neq 0$, wherein the standard minimal
couplings mediated via the Lorentz connection are \emph{subleading} at energy
scales $\sqrt{|\L|}<\!\!\!< E<\!\!\!< M_{\rm p}$.
Finally, combining our results with those obtained by Metsaev,
we give the complete list of \emph{all} the manifestly covariant
cubic couplings of the form $1-s-s\,$ and $2-s-s\,$,
in Minkowski background.
\end{abstract}

\vspace*{.2cm}

\setcounter{page}{1}

\pagebreak

\tableofcontents

\section{Introduction and overview}

\subsection{No-go and yes-go results for $\L=0$}

From a general perspective it is a remarkable fact that the full gravitational
couplings of lower-spin fields involve at most two derivatives in the
Lagrangian. For spin $s\leqslant 1$ the \emph{standard} covariantization
scheme, wherein $\partial \rightarrow \nabla = \partial + \o \,$ with $\o$ being a
torsion-constrained Lorentz connection, induces the ``minimal coupling''
$\int d^Dx\, h_{\mu\nu} T^{\mu\nu}$ where $T^{\mu\nu}$ is the
Belifante-Rosenfeld stress-tensor which is quadratic and contains up to two
derivatives. Actually, for scalars, Maxwell fields and other Lorentz-invariant
differential forms, the Lorentz covariantization is trivial and the coupling
therefore involves no derivatives of the metric. It is also remarkable that
the non-abelian cubic self-coupling of a spin-2 field contains only two
derivatives.

Turning to gauge fields with $s>2$ and considering $2-s-s\,$ couplings in
an expansion around flat spacetime, the standard scheme breaks down as has
been known for a long time \cite{Aragone:1979bm,Berends:1979wu,Aragone:1981yn}.
These no-go results have recently been strengthened in
\cite{Metsaev:2005ar,Metsaev:2007rn} following a light-cone
method and in \cite{Porrati:2008rm} with $S$-matrix tools.
More interestingly, in the works \cite{Metsaev:2005ar,Boulanger:2006gr,Metsaev:2007rn}
some \emph{yes-go} results have been obtained.
In the specific case of $s=3\,$, the work \cite{Boulanger:2006gr}
provides a manifestly covariant \emph{non-standard} four-derivative vertex associated with a \emph{nonabelian} deformation of the gauge algebra.
These yes-go results suggest a class of minimal nonabelian\footnote{We consider only
couplings that truly deform the initial abelian gauge algebra into a
nonabelian one, similarly to what happens when coupling $N^2-1$ Maxwell
fields in order to obtain the Yang-Mills $SU(N)$ theory.
Interesting results and references on abelian couplings can be found in the
review \cite{Fotopoulos:2008ka}. } non-standard vertices containing $2s-2$ derivatives.
We wish to emphasize that the existence of cubic couplings containing $2s-2$ derivatives
was explicitly shown in \cite{Metsaev:2005ar,Metsaev:2007rn}
although the light-cone gauge method used therein does not
exhibit the nature of the gauge algebra and does not
readily allow for the explicit construction of the corresponding
covariant vertices. The results are nonetheless remarkable in that
they show the existence of only a few non-trivial cubic vertices
of the general form $s-s'-s''\,$ for massive and massless
fields (bosonic and fermionic) in flat space of arbitrary dimension
$D>3\,$. In the case of integer spins, the possible vertices have $s+s'+s''-p$ derivatives where $p=0,2,\dots,2\min(s,s',s'')$.

In the specific massless $2-3-3\,$ case, using the BRST-BV cohomological
methods of \cite{Barnich:1993vg,Henneaux:1997bm},
the vertex of \cite{Boulanger:2006gr} was shown to be
\emph{unique} among the class of vertices that:
{(i)} contain a \emph{finite} number of derivatives;
(ii) manifestly preserve Poincar\'e invariance and
{(iii)} induce a \emph{nonabelian} deformation of the gauge algebra.
This uniqueness result relies on the fact that other candidate nonabelian
deformations cannot be ``integrated'' cohomologically to gauge
transformations and vertices.
We have managed to push the uniqueness analysis to the case of $s=4$ and
the unique $2-4-4\;$ nonabelian vertex is presented
in Section \ref{sec:244} together with
its corresponding gauge algebra and transformations.

We also extend the results of~\cite{Boulanger:2006gr} with
the cohomological proof in Section \ref{sec:2ss}
that the standard two-derivative minimal couplings $2-s-s\,$ are
inconsistent, thereby providing an alternative
proof for the results recently obtained in
\cite{Metsaev:2005ar,Metsaev:2007rn} following light-cone methods and in \cite{Porrati:2008rm}
following $S$-matrix methods.
In the same section \ref{sec:2ss},
combining the cohomological approach with the light-cone
results of Metsaev~\cite{Metsaev:2005ar,Metsaev:2007rn},
we show that there exists \emph{only one} nonabelian $2-s-s\,$
coupling, which contains $2s-2$ derivatives and must be the flat limit of the well-known nonabelian Fradkin--Vasiliev vertex~\cite{Fradkin:1986qy,Fradkin:1987ks}
in $AdS\,$, as we verify explicitly for $s=3$. There also exist two abelian covariant $2-s-s\,$ vertices containing $2s+2$ and $2s$ derivatives. Their existence was first found in \cite{Metsaev:2005ar}, and we exhibit them here explicitly in their covariant form. The $(2s+2)$-derivative vertex is of the Born-Infeld type, whereas the $2s$-derivative vertex exists only for $D\geqslant 5$ and is gauge invariant up to a total derivative.
These three vertices, with $2s-2$, $2s$ and $2s+2$ derivatives, thus exhaust the possibilities of manifestly Lorentz-covariant $2-s-s\,$ couplings in flat space.

We begin in Section \ref{sec:1ss} by examining the simpler case of $1-s-s\,$ vertices. We build explicitly the unique, nonabelian
$1-s-s\,$ coupling, which has $2s-1$ derivatives, together with the only abelian
$1-s-s\,$ vertex, which as $2s+1$ derivatives, thereby completing the list of all possible nontrivial, manifestly covariant, $1-s-s\,$ couplings.
Again, by the uniqueness of the nonabelian vertex, we know
that it is the flat limit of the corresponding
$AdS$ Fradkin--Vasiliev (FV) vertex~\cite{Fradkin:1986qy,Fradkin:1987ks}.

\subsection{The Fradkin--Vasiliev cancelation mechanism for $\L\neq 0$}

Under the assumptions that the cosmological constant vanishes and that the
Lagrangian contains at most two derivatives, the standard covariantization of
Fronsdal's action leads to an inconsistent cubic action of the
form\footnote{We use mostly positive signature and $R=g^{\m\r} g^{\n\s}
R_{\m\n\r\s}$. The Fierz-Pauli action $\int d^Dx(-\frac12 \partial^\m
h^{\r\s}\partial_\m h_{\r\s}+\cdots)$ is recuperated modulo boundary terms
from $\frac{1}{(\ell_p)^{D-2}}\int d^D x\sqrt{-g}R(g)$ upon substituting
$g_{\mu\nu}=\eta_{\mu\nu}+\sqrt{2}(\ell_p)^{\frac{D-2}2}h_{\mu\nu}$.}
\begin{eqnarray}
  S^{\L=0}_{2ss}[g,\phi] &=& {1\over \ell_p^{D-2}}\int
\left(R+G+\frac12 W_{\m\n\r\s}\b_{(2)}^{\m\n,\r\s}(\phi^{\otimes 2})\right)\ ,
\end{eqnarray}
where $\ell_p$ is the Planck length, $\int=\int d^Dx \sqrt{-g}$, the spin-$s$
kinetic term\footnote{The initial choice of free kinetic terms affects the
classical anomaly and the final form of anomaly cancelation terms.}
$G=\frac12 \phi^{\mu(s)}G_{\mu(s)}(g;\nabla\nabla\phi)$ with the Einstein-like
self-adjoint operator\footnote{Repeated indices distinguished by
sub-indexation are implicitly symmetrized, $\nabla\cdot V_{\m(s-1)}\equiv
\nabla^\n V_{\n\m(s-1)}$ and $V'_{\mu(s-2)}\equiv g^{\n\r}V_{\n\r\m(s-2)}$.}
\begin{eqnarray}
G_{\mu(s)}&=&F_{\mu(s)}-\frac{s(s-1)}4 g_{\m(2)} F'_{\m(s-2)}\ ,\\[5pt]
F_{\m(s)}&=& \nabla^2\phi_{\m(s)}-s\nabla_{\m_1}\nabla\cdot
\phi_{\m(s-1)}+\frac{s(s-1)}2\nabla_{\m_1} \nabla_{\m_2}\phi'_{\m(s-2)}\ ,
\end{eqnarray}
the covariantized Fronsdal field strength. The symbol $\b_{(2)}$ denotes a
dimensionless symmetric bilinear form, $W_{\m\n\r\s}$ is the spin-2 Weyl
tensor, and $(\ell_p)^2W$ and $\phi$ are assumed to be weak fields. A quantity
${\cal O}$ has a regular weak-field expansion if
${\cal O}=\sum_{n=n({\cal O})}^\infty \stackrel{(n)}{{\cal O}}$ where
$\stackrel{(n)}{{\cal O}}$ scales like $g^{n}$ if the weak fields are rescaled
by a constant factor $g$, and we shall refer to $\stackrel{(n)}{{\cal O}}$ as
being of $n$th in weak fields, or equivalently, as being of order
$n-n({\cal O})$ in the $g$ expansion. Under the spin-$s$ gauge transformation
$\d_\e\phi_{\mu(s)} =s\,\nabla_{\m_1}\e_{\mu(s-1)}+R_{\m(s)}[g_{\a\b},\phi,
\e]$ and $\d_\e g_{\m\n}=R_{\m\n}[g_{\a\b},\phi,\e]$, where $\e$ is a weak
traceless parameter and $R_{\m(s)}$ and $R_{\m(2)}$ are quadratic in weak
fields, the variation of the action picks up the first-order contribution
\begin{eqnarray}
  \delta_\e \int G&=& \int W^{\m\n\r\s}{\cal A}_{\m\n,\r\s}(g_{\a\b};
\nabla\phi\otimes\e)\ ,
\end{eqnarray}
where the bilinear form
\begin{eqnarray}
{\cal A}_{\m\n,\r\s}&=& 2s(s-1){\bf P}^W
\Big[\nabla_\m \phi_{\n\s\t(s-2)}\e_{\r}{}^{\t(s-2)} \nonumber\\[5pt]&&
+(s-2)\big(\nabla_\s\phi'_{\n\t(s-3)}-\frac12 \,
\nabla\cdot \phi_{\n\s\t(s-3)}+\frac{(s-3)}4
\nabla_{\t_1}\phi'_{\n\s\t(s-4)}\big)\e_{\m\r}{}^{\t(s-3)}\Big]\ ,
\end{eqnarray}
that has been shown to be anomalous for $s=3$ \cite{Aragone:1981yn} (for
recent re-analysis see \cite{Boulanger:2006gr} and also \cite{Porrati:2008rm}
for an $S$-matrix argument) in the sense that it cannot be canceled by any
choice of $\b_{(2)}$ nor by abandoning the assumption that the Lagrangian
contains at most two derivatives.

However, as first realized by Fradkin and Vasiliev \cite{Fradkin:1986qy}, if
both $\L\neq 0$ and higher-derivative terms are added to the cubic part of the
action, the analogous obstruction can be bypassed. In the weak-field expansion
the resulting \emph{minimal cubic action} reads
\begin{eqnarray}
  S^{\L}_{2ss}[g,\phi] &=&
  \frac1{\ell_p^{D-2}}\int \left(R(g)-\L+G_\L\right)+
  \sum_{ \substack{n=2 \\ {n~even}}}^{n_{\rm min}(s)}
  \frac1{\ell_p^{D-2}}\int V^{(n)}_\L(2,s,s)\ ,\\[5pt]
  V^{(n)}_\L(2,s,s)&=& \frac1{2\l^{n-2}} \sum_{p+q=n-2}
  W_{\m\n\r\s}\b^{\m\n,\r\s}_{(n);p,q}(\nabla^p\phi\otimes \nabla^q\phi)\ ,
\end{eqnarray}
with $\l^2\equiv -\frac{\L}{(D-1)(D-2)}$ and $G_\L$ is the Einstein-like
kinetic term built from $F_\L=F-\frac12 \l^2 M_s^2(\phi^{\otimes 2})$ (see
(\ref{Fs}) below). The spin-$s$ gauge invariance up to first order uses that
at zeroth order
\begin{eqnarray}
  &&[\nabla_\mu,\nabla_\nu] V_\r \ =\ R_{\mu\nu\r}{}^\s V_\s\ \approx\
2\l^2g_{\r[\n} V_{\m]}+W_{\m\n\r}{}^\s V_\s\ ,\qquad \nabla^\mu W_{\m\n,\r\s}\
\approx 0\ ,\label{nablanabla}\\[5pt]
  &&R_{\m\n}-\frac12(R-\L) g_{\m\nu}\ \approx\ 0\ ,\qquad F_{\m(s)}-
\frac12\, \l^2 M_s^2(\phi^{\otimes 2})\ \approx\ 0\ ,
\end{eqnarray}
where $\approx$ is used for equalities that hold on-shell. At zeroth order,
the invariance requires the critical mass matrix \cite{Fronsdal:1978vb}
\begin{eqnarray}
 M_s^2(\phi^{\otimes 2})&=& m_s^2\phi^2+m^{\prime 2}_s\phi^{\prime 2}\ ,\quad
m_s^2\ =\ s^2+(D-6)s-2D+6\ ,\quad m^{\prime 2}_s\ =\ s(s-1)\ .
\end{eqnarray}
At first order, the classical anomaly $\int W{\cal A}$, which is independent
of $\l$, is accompanied by two types of $\l$-independent counter terms, namely
$\d_\e \int V^{(2)}_\L$ plus the contributions to $\delta_\e \int V^{(4)}_\L$
from the constant-curvature part of $[\nabla,\nabla]$, that can be arranged to
cancel the anomaly at order $\l^0$. At order $\l^{-2}$, the remaining terms in
$\delta_\e \int V^{(4)}_\L$ can be canceled against order $\l^{-2}$
contributions from $\delta_\e \int V^{(6)}_\L$ and so on, until the procedure
terminates at the \emph{top vertex}
$V^{{\rm top}}_\L(2,s,s)=V^{(n_{\rm min}(s))}_\L$ that: {(i)} is
\emph{weakly gauge invariant} up to total derivatives and terms that are of
lower order in $\l$; and {(ii)} contains a total number of derivatives given
by
\begin{eqnarray}
n_{\rm min}(s)=2s-2\ .
\end{eqnarray}
Counting numbers of derivatives, there is a \emph{gap} between the top vertex
and the \emph{tail} of Born-Infeld-like non-minimal cubic vertices,
which is \emph{a priori} of the form
\begin{eqnarray}
S^{\rm nm}_{2ss;\L}&=& \sum_{n=0}^\infty
\frac1{(\ell_p)^{D-2}2\l^{2(n+s)}}\sum_{p+q=2n}\int W_{\m\n\r\s}
\c^{\m\b,\r\s}_{(n);p,q}(\nabla^pC\otimes \nabla^q C)\ ,\label{BItail}
\end{eqnarray}
where $C_{\m(s),\n(s)}$ is the linearized spin-$s$ Weyl tensor and $\c_{(n);p,q}$ are
dimensionless bilinear forms. Adapting the flat-space result of \cite{Metsaev:2005ar} to constantly curved backgrounds suggests that, if the $\c_{(n);p,q}$ fall off with $n$ sufficiently fast, then the couplings with $n\geqslant 1$ can be removed by a suitable, possibly non-local, field redefinition. More generally, turning to higher orders in the weak-field expansion, one may adopt the \emph{canonical} frame of standard fields that by definition minimizes the maximal numbers of derivatives at each order.

The \emph{existence} of at least one cancelation procedure has sofar been shown in the literature only for $D=4,5$ \cite{Fradkin:1986qy,Vasiliev:2001wa,Alkalaev:2002rq}, following the existence of a more general minimal cubic action given within the frame-like
formulation based on a nonabelian higher-spin Lie algebra extension
$\mathfrak{h}$ of $\mathfrak{so}(D+1;\mathbb{C})$. The 4D action is a natural
generalization of the MacDowell-Mansouri action for $\L$-gravity. It is given
by a four-form Lagrangian based on a bilinear form
$<\cdot,\cdot>_{\mathfrak{h}}$ such that the resulting action: {(i)} contains
at most $2$ derivatives at second order in weak fields; {(ii)} propagates
symmetric rank-$s$ tensor gauge fields with $s\geqslant 1$ and critical mass;
{(iii)} contains nonabelian $V^{(n)}_{\L}(s,s',s'')$ vertices with
$s,s',s''\geqslant 1$ and $n\leqslant n_{\rm min}(s,s',s'')$. The 5D action shares the
same basic features \cite{Vasiliev:2001wa,Alkalaev:2002rq}. The existence
issue in $D>5$ is open at present though all indications sofar hint at that
the lower-dimensional cases do actually have a generalization to arbitrary
$D$.

\subsection{Recovering the metric-like FV 2-3-3 vertex}\label{FVsection}

Apparently Fradkin and Vasiliev first found the gravitational coupling of the
spin-3 field using the metric-like formalism without publishing their result
(see \cite{Vasiliev:2001ur} for an account). Later they obtained and published
their (by now famous) result in the frame-like formalism in the general
$2-s-s$ case in $D=4$ \cite{Fradkin:1986qy}. For the purpose of discussing the
uniqueness of their result and its extension to $D$ dimensions, we need the
explicit form of the $D$-dimensional $2-3-3\,$~ FV vertex. To this end, we
work within the metric-like formulation and start from the free Lagrangian
${\cal L}_{2}+{\cal L}_{3}\,$
where Fronsdal's Lagrangian for a symmetric rank-$s$ tensor gauge field in
$AdS_D$ reads \cite{Buchbinder:2001bs}
\begin{eqnarray}-\frac{\cl_{s}}{\sqrt{-\bar g}} &=&\frac{1}{2}\,
\overline\N_\m \phi_{\a_1...\a_s}\overline\N^\m \phi^{\a_1...\a_s} -
\frac{1}{2}\,s\overline\N^\m \phi_{\m\a_1...\a_{s-1}}\overline\N_\n
\phi^{\n\a_1...\a_{s-1}}\nonumber\\&& + \frac{1}{2}\,s(s-1)\overline\N_\a
\phi'_{\b_1...\b_{s-2}}\overline\N_\m \phi^{\m\a\b_1...\b_{s-2}}
 - \frac{1}{4}\,s(s-1)\overline\N_\m \phi'_{\a_1...\a_{s-2}}\overline\N^\m
\phi'^{\a_1...\a_{s-2}}\nonumber\\&& -\frac{1}{8}s(s-1)(s-2)\overline\N^\m
\phi'_{\m\a_1...\a_{s-3}}\overline\N_\n \phi'^{\n\a_1...\a_{s-3}}
\nonumber\\&&
+\frac12 \,\l^2\left[s^2+(D-6)s-2D+6\right]\phi_{\a_1...\a_s}
\phi^{\a_1...\a_s}\nonumber\\&&-\frac14 \l^2 s(s-1)
\left[s^2+(D-4)s-D+1\right]\phi'_{\a_1...\a_{s-2}}\phi'^{\a_1...\a_{s-2}}\ ,
\label{Fs}\end{eqnarray}
given that
$\overline R_{\a\b\c\d}=-\l^2(\bar g_{\a\g}\bar g_{\b\d}-\bar g_{\b\g}\bar
g_{\a\d})$. We find, using the Mathematica package Ricci
\cite{Lee}, that the 2-3-3 FV vertex is given by\footnote{We use conventions
where $h_{\a\b}$ and $\phi_{\a\b\c}$ are dimensionless. The linearized spin-2
Weyl tensor $w_{\a\b\c\d}=s_{\a\b\c\d}-\frac{2}{D-2}(\bar
g_{\a[\g}s_{\d]\b}-\bar g_{\b[\g}s_{\d]\a})
+\frac{2}{(D-1)(D-2)}\bar g_{\a[\g}\bar g_{\d]\b}s$, where
$s_{\a\b\c\d}\equiv-\overline \nabla_{\g}\overline
\nabla_{[\a}h_{\b]\d}+\overline \nabla_{\d}\overline \nabla_{[\a}h_{\b]\d}+
\l^2(\bar g_{\g[\a}h_{\b]\d}-\bar g_{\d[\a}h_{\b]\g})\,$ has the property that
at zeroth order $\bar g^{\a\b}s_{\a\c\b\d}\approx0$ and $\overline
\nabla^{\a}s_{\a\b\c\d}\approx 0$. The form of the 2-3-3 FV vertex given in
(\ref{FVvertex}) reflects the initial choice of free Lagrangian made in
(\ref{Fs}).}
\begin{eqnarray}-{\stackrel{(3)}{\cl}_{FV}\over
\sqrt{-\bar g}}&\approx&-\frac{11}{2}~
w_{\a\b\c\d}~\phi^{\a\c}_{\phantom{\a\c}\m}\phi^{\b\d\m}
+ \frac{1}{(D-1)\l^2} ~w_{\a\b\c\d}~\Big[2\ \phi'_\m
\overline \N^{(\b}\overline \N^{\d)}\phi^{\a\c\m}
+ \phi^{\a\c}_{\phantom{\a\c}\m}\overline \N^{(\d}\overline
\N^{\m)}\phi'^{\b}\nonumber\\[5pt]
&&- 3\
\phi'^{\a}\overline \N^{(\d}\overline
\N^{\m)}\phi^{\b\c\phantom{\m}}_{\phantom{\b\c}\m}
+ 2\ \phi^{\a}_{\phantom{\a}\m\n}\overline \N^{(\d}\overline
\N^{\n)}\phi^{\b\g\m}
+ \overline \N_\m\phi^{\a\c\m}_{\phantom{\a\c\m}}\overline
\N_\n\phi^{\b\d\n}_{\phantom{\b\d\n}}
- \phi^{\a\c\m}\overline \N_{(\m}\overline
\N_{\n)}\phi^{\b\d\n}_{\phantom{\b\d\n}}\nonumber\\[5pt]&&
- 2\ \overline \N^{(\m}\phi^{\n)\a\c}\overline
\N^{\phantom{\m}}_\m\phi^{\b\d}_{\phantom{\b\d}\n}
- 2\
\phi^{\a\c}_{\phantom{\a\c}\m}\overline \N^{(\d}\overline \N^{\n)}
\phi^{\b\m\phantom{\n}}_{\phantom{\b\m}\n}
+ \phi'^{\a}\overline \N^{(\b}\overline \N^{\d)}\phi'^{\g}
- \phi^{\a}_{\phantom{\a}\m\n}\overline \N^{(\b}\overline
\N^{\d)}\phi^{\c\m\n}\Big]\;.\qquad
\label{FVvertex}\end{eqnarray}
The first term corrects the obstruction to the standard minimal scheme
at the expense of introducing a new one that can be removed, however, by
adding the above particular combination of two-derivative terms (involving
only a subset of all possible tensorial structures as expected from the frame-
like formulation). We stress again that the top vertex
does not introduce any further obstructions, and that the vertex
indeed exhibits the gap.

\subsection{Non-Uniform $\L\rightarrow 0$ Limits}

Since for given $s$ the derivative expansion of the minimal $2-s-s$ coupling
terminates at the top vertex $V^{(2s-2)}_\L(2,s,s)$, the cubic action
$S^{\L}_{2ss}$ admits the \emph{scaling limit}
\begin{eqnarray}
\l&=&\e (\ell_p)^{-1} \ ,\quad W\ =\ \e^{2s-4}\widetilde{W}\ ,\qquad \e\
\rightarrow\ 0\ ,
\end{eqnarray}
with evanescent piece $\widetilde W_{\m\n\r\s}$ held fixed, so that
$\widetilde W_{\m\n\r\s}$ can be replaced by the linearized Weyl tensor
$\widetilde w_{\m\n\r\s}$ in the cubic vertices, resulting in the action
\begin{eqnarray}
\widetilde S^{\L=0}_{2ss}[g,\phi]&=&{1\over \ell_p^{D-2}}\int d^Dx \sqrt{-g}
\left(R(g)+G_0+\widetilde V^{(2s-2)}_0(2,s,s)\right)\ ,\\[5pt]
\widetilde V^{(2s-2)}_0(2,s,s)&=&\frac12 \sum_{p+q=2s-4}\int \widetilde
w_{\m\n\r\s}\b^{\m\n,\r\s}_{(n);p,q}(\nabla^p\phi\otimes \nabla^q\phi)\ ,
\end{eqnarray}
that is faithful up to cubic order in weak graviton and spin-s fields, and
$G_0$ contains the connection $\nabla_0$ obeying the flatness condition
$[\nabla_0,\nabla_0]=0$.

Alternatively, one may first perturbatively expand the FV action around AdS
and then take the $\L\rightarrow 0$ limit as follows:
\begin{eqnarray}
\l&=&\e\widetilde\ell_p^{-1}\ ,\qquad \ell_p\ =\ \e^{\D_p}\widetilde \ell_p\ ,
\\[5pt] h_{\m\n}&=&\e^{\D_h}\widetilde h_{\m\n}\ ,\qquad\phi_{\m\n\r}\ =\
\e^{\D_\phi}\widetilde\phi_{\m\n\r}\ ,\qquad \e\ \rightarrow\ 0\ ,
\end{eqnarray}
with $\widetilde\ell_p$, $\widetilde h$ and $\widetilde \phi$ kept fixed and
$\D_h=\D_\phi=2(s-2)$ and $\D_p=\frac{4(s-2)}{D-2}$. The resulting flat-space
2-3-3 vertex reads
\begin{eqnarray}-\stackrel{(3)}{\cl}&=&\frac{1}
{D-1}\tilde{w}_{\a\b\c\d}\Big[2\
 \widetilde\phi'_\m \6^{\b}\6^{\d}\widetilde\phi^{\a\c\m}
+ \widetilde\phi^{\a\c}_{\phantom{\a\c}\m}\6^{\d}\6^{\m}\widetilde\phi'^{\b}
- 3\ \widetilde\phi'^{\a}\6^{\d}\6^{\m}\widetilde
\phi^{\b\c\phantom{\m}}_{\phantom{\b\c}\m}
\nonumber\\[5pt]&&
+ 2\ \widetilde\phi^{\a}_{\phantom{\a}\m\n}\6^{(\d}\6^{\n)}\widetilde
\phi^{\b\g\m}+ \6_\m\widetilde\phi^{\a\c\m}_{\phantom{\a\c\m}}
\6_\n\widetilde\phi^{\b\d\n}_{\phantom{\b\d\n}} -
\widetilde\phi^{\a\c\m}\6_{\m}\6_{\n}\widetilde
\phi^{\b\d\n}_{\phantom{\b\d\n}}\nonumber\\[5pt]&&
- 2\
\6^{(\m}\widetilde\phi^{\n)\a\c}\6^{\phantom{\m}}_\m
\widetilde\phi^{\b\d}_{\phantom{\b\d}\n}- 2\
\widetilde\phi^{\a\c}_{\phantom{\a\c}\m}\6^{\d}\6^{\n}
\widetilde\phi^{\b\m\phantom{\n}}_{\phantom{\b\m}\n}
+ \widetilde\phi'^{\a}\6^{\b}\6^{\d}\widetilde\phi'^{\g}
- \widetilde\phi^{\a}_{\phantom{\a}\m\n}\6^{\b}\6^{\d}
\widetilde\phi^{\c\m\n}\Big
]\end{eqnarray}
where $\tilde{w}_{\a\b\c\d}=\tilde{K}_{\a\b\c\d}-\frac{2}{D-2}\,
(\eta_{\a[\g}\tilde{K}_{\d]\b}-\eta_{\b[\g}\tilde{K}_{\d]\a})+
\frac{2}{(D-1)(D-2)}\,\eta_{\a[\g}\eta_{\d]\b}\tilde{K}$ with
$\tilde{K}_{\a\b\c\d}=-\6_{\g}\6_{[\a}\tilde{h}_{\b]\d}
+\6_{\d}\6_{[\a}\tilde{h}_{\b]\d}$.

As discussed above, the top 2-3-3 vertex must be equivalent modulo total
derivatives and linearized equations of motion to the nonabelian 2-3-3 vertex
presented in Appendix B of \cite{Boulanger:2006gr} which we have verified
explicitly\footnote{Modulo the Bianchi identities of $R_{\mu\nu}$, there are
49 four-derivative terms that are proportional to the spin-2 field equations:
25 terms of the form $R_{..}\6^2_{..}\phi_{...}\phi_{...}$ and 24 terms of the
form
$R_{..}\6_{.}\phi_{...}\6_{.}\phi_{...}$. Adding an arbitrary linear
combination of these to the vertex, lifting the derivatives from $\tilde{h}$,
subtracting the ``cohomological'' vertex, and finally factoring out the spin-3
equations of motion by eliminating $\partial^2\phi$, yields a
simple system of equations that allow us to fit the coefficients.}.

\subsection{Uniqueness of the $2-3-3\,$ FV vertex}

The uniqueness of the FV cancelation procedure in the case of spin $s=3$ can
be now be established for any $D$ as follows. We obtained the $AdS_D$
covariantization
$$S^{\L}[h,\phi]=S^{\Lambda}_{free} + g\,S^{\Lambda}_{cubic}$$
of the nonabelian flat spacetime action
$$S^{\L=0}[h,\phi]=S^{Flat}_{free} + g\,S^{Flat}_{cubic}$$
obtained in \cite{Boulanger:2006gr},
with $S^{\L=0}_{cubic}= \int d^Dx\, V^{(4)}_0(2,3,3)\,$ and
$g$ the deformation parameter. The cubic part
$S^{\Lambda}_{cubic}=\int d^Dx \sqrt{-g}\; V_{\Lambda}(2,3,3)$
possesses an expansion in powers of the $AdS$ radius, where the
contribution to $V_{\Lambda}(2,3,3)$ with the maximum number of derivatives
is called $V^{top}_{\Lambda}(2,3,3)\,$.
We recall that, using the power of the BRST-BV cohomological
method \cite{Barnich:1993vg}, the first-order deformation $S^{\L=0}[h,\phi]$
has been proved \cite{Boulanger:2006gr} to be \emph{unique} under
the sole assumptions of
\begin{itemize}
\item Locality,
\item Manifest Poincar\'e invariance,
\item Nonabelian nature of the deformed gauge algebra.
\end{itemize}
The last assumption allows the addition of Born-Infeld-like cubic vertices of
the form $V^{BI}_0(2,3,3)=C(h)C(\phi)C(\phi)\,$ where
$C(h)$ and $C(\phi)$ denote linearized Weyl tensors and we note that $C(\phi)$
contains $3$ derivatives \cite{Damour:1987vm}. Such vertices are strictly
gauge invariant and do not deform the gauge algebra nor the transformations.
We also disregard deformations of
the transformations that do not induce nonabelian gauge algebras, as is the
case for such deformations involving the curvature tensors.
In the following, when we refer to a deformation as unique it should be
understood to be up to the addition of other deformations that do not deform
the gauge algebra.

The uniqueness of $S^{\L=0}[h,\phi]$ is instrumental in showing the
uniqueness of its $AdS_D$ completion $S^{\L}[h,\phi]\,$,
due to the linearity of the perturbative deformation scheme and the smoothness
of the flat limit at the level of cubic actions. The proof goes as follows.
First suppose that there exists another action
$S^{\prime\L}[h,\phi]=S^{\Lambda}_{free} + g\,S^{\prime\Lambda}_{cubic}$
that admits a nonabelian gauge algebra and whose top vertex
$V_{\Lambda}^{\prime {\rm top}}(2,3,3)$ involves $n_{\rm top}$ derivatives
with $n_{\rm top}\neq 4$. Then, this action would scale to a nonabelian
flat-space action whose cubic vertex would involve $n_{\rm top}$ derivatives.
This is impossible, however, because the \emph{only} nonabelian cubic vertex
in flat space is $V^{(4)}_0(2,3,3)\,$.
Secondly, suppose there exists a nonabelian action
$S^{\prime\prime\L}[h,\phi]=S^{\Lambda}_{free}
+g\,S^{\prime\prime\Lambda}_{cubic}$
whose top vertex contains $4$ derivatives but is otherwise different from
$V_{\Lambda}^{\rm top}(2,3,3)$.
Then its flat limit would yield a theory with a cubic vertex, involving
4 derivatives, but different from $V^{\rm top}_{0}(2,3,3)\,$, which is
impossible due to the uniqueness of the latter deformation.
Thirdly, and finally, suppose there exists a cubic action with top vertex
$V^{\rm top}_{\Lambda}(2,3,3)\,$ but differing from $S^{\L}_{cubic}$
in the vertices with lesser numbers of derivatives.
By the linearity of the BRST-BV deformation scheme, the difference between
this coupling and $S^{\L}_{cubic}$ would lead to a nonabelian theory in $AdS$
with top vertex involving less than 4 derivatives.
Its flat-space limit would therefore yield a nonabelian action
whose top vertex would possess less than 4 derivatives, which is impossible
due to the uniqueness of $S^{\L=0}[h,\phi]\,$.

A more rigorous proof can be stated entirely in terms of master actions within
the BRST-BV framework.
Then all ambiguities resulting from trivial field and gauge parameter
redefintions are automatically dealt with cohomologically. Moreover, the
possibility of scaling away the nonabelianess while at the same time retaining
the vertex is ruled out\footnote{Consider a master action
$W_\l=\stackrel{(0)}{W_\l}+g\stackrel{(1)}{W_\l}+\cdots$ with
$\stackrel{(1)}{W_\l}=\int (a^\l_2+a^\l_1+a^\l_0)$ where $a^\l_2$, $a^\l_1$
and $a^\l_0$, respectively, contain the nonabelian deformation of the gauge
algebra, the corresponding gauge transformations and vertices. The master
equation amounts to $\c^\l a^\l_2=0$, $\c^\l a_1^\l+\d^\l a_2^\l=d c_1^\l$ and
$\c^\l a_0^\l+\d^\l a^\l_1=d c_0^\l$ where $\c^\l$ and $\d^\l$ have $\l$
expansions starting at order $\l^0$. Since the system is linear and determines
$a_1^\l$ and $a_0^\l$ for given $a_2^\l$ it follows that all $a_i^\l$ scale
with $\l$ the same way in the limit $\l\rightarrow 0$.}.

\subsection{On Separation of Scales in Higher-Spin Gauge Theory}

Thanks to Vasiliev's oscillator constructions \cite{Vasiliev:1990en,Vasiliev:2003ev} it has been established that fully nonlinear nonabelian higher-spin gauge field equations exist in arbitrary dimensions in the case of symmetric rank-$s$ tensor gauge fields.
Compared to the cubic actions, the full equations exhibit two additional essential features: {(i)} a precise spectrum $\mathfrak{D}$ given by an
\emph{infinite tower} of $\mathfrak{so}(D+1;\mathbb{C})$ representations
forming a unitary representation of (a real form of) the higher-spin algebra $\mathfrak{h}$ (see \emph{e.g.} \cite{Konstein:1989ij}); {(ii)} nonlocal, potentially \emph{infinite}, Born-Infeld tails.

The closed form of Vasiliev's equations requires the \emph{unfolded formulation} of field theory
whereby \cite{D'Auria:1982my,D'Auria:1982nx,van Nieuwenhuizen:1982zf,Vasiliev:1988xc,Vasiliev:1988sa,Vasiliev:2001ur,Skvortsov:2008vs}: {(i)} standard physical (gauge) fields are replaced as independent action variables by differential forms taking their values in $\mathfrak{so}(D+1;\mathbb{C})$ modules that are finite-dimensional for $p$-forms with $p>0$ and infinite-dimensional for zero-forms; {(ii)} the resulting kinetic terms feature only the exterior derivative $d$; {(iii)} the standard interactions are mapped to non-linear structure functions appearing in the unfolded first-order equations obeying \emph{algebraic conditions} assuring $d{}^2=0$.
Thus the on-shell content of a spin-$s$ gauge field $\phi_s$ is mapped into an
infinite-dimensional collection of zero-forms carrying traceless
Lorentz indices filling out the covariant Taylor expansion on-shell of the corresponding
Weyl tensor $C(\phi_s)$. Letting $X^\a$ denote the complete unfolded field
content, the unfolded equations take the form $dX^\a+f^\a(X^\b)=0$ where $f^\a$ are written entirely using exterior algebra, and subject to the algebraic condition $f^\b{\partial^l\over\partial X^\b} f^\a=0\,$ (defining what is sometimes referred to as a free-differential algebra).
The salient feature of the unfolded framework is that any consistent
deformation is automatically gauge-invariant in the sense that every $p$-form
with $p\geqslant 1$ is accompanied by a $(p-1)$-form gauge parameter, independently of
whether the symmetry is manifestly realized or not.

Vasiliev's equations provide one solution to the on-shell deformation problem given a one-form $A$ taking its values in the algebra $\mathfrak{h}$, and a zero-form $\Phi$ containing all Weyl tensors and their on-shell derivatives, which is the unfolded counterpart of the massless representation $\mathfrak{D}$. The embedding of the canonical fields $\{g_{\m\n},\phi,\dots\}$ into $\Phi$ and $A$ requires a non-local field redefinition\footnote{The situation in higher-spin gauge theory is analogous to that in string theory: in both cases the microscopic formulation is defined in terms of ``vertex operators'' living in an associative algebra associated with an ``internal'' quantum theory. As a result, the graviton vertex receives corrections leading to a microscopic frame that is different from the canonical Einstein frame (see \cite{Sezgin:2005hf} for a related discussion).} to microscopic counterparts $\{\widehat g_{\m\n},\widehat \phi,\dots\}$. In the microscopic frame, the standard field equations are non-canonical and actually contain infinite Born-Infeld tails already at first order in the weak-field expansion (see \cite{Sezgin:2002ru} for a discussion). For example, the first-order corrections to the stress tensor, defined by $\widehat R_{\m\n}-\frac12 \widehat g_{\m\n}(\widehat R-\L)=\widehat T_{\m\n}$, from a given spin $s$ arise in a derivative expansion of the form $\widehat T^{(1)}_{\mu\nu}=\sum_{n=0}^\infty
\sum_{p+q=2n}\l^{-2n}\widehat T^{(n);p,q}_{\mu\nu}(\widehat \nabla^p\widehat \phi_s,
\widehat \nabla^q\widehat \phi_s)$ where $\widehat \nabla^p\widehat \phi_s$ is a connection if $p<s$ and $(p-s)$ derivatives of $C(\widehat \phi_s)$ if $p\geqslant s$ (see \emph{e.g.} \cite{Kristiansson:2003xx} for the case of $s=0$).

As discussed below (\ref{BItail}), the microscopic tails should be related to the canonical vertices via non-local, potentially divergent, field redefinitions. Thus one has the
following scheme:
\begin{eqnarray}
\begin{array}{c}\mbox{Unfolded}\\[-3pt]\mbox{master-field}\\[-3pt]
\mbox{equations}\end{array}\quad& \stackrel{\tiny\begin{array}{c}{\rm weak}\\{\rm
fields}\end{array}}{\leftrightarrows} &\quad \begin{array}{c}\mbox{Standard-exotic}\\
[-3pt]\mbox{microscopic}\\[-3pt]\mbox{field equations}\end{array}\qquad \stackrel{\tiny\begin{array}{c}{\rm non-local}\\
{\rm field\, redef.}\end{array}}{\rightleftarrows}\qquad \begin{array}{c}\mbox{Standard-exotic}\\
[-3pt]\mbox{canonical}\\[-3pt]\mbox{field equations}\end{array}
\end{eqnarray}
Thus, the semi-classical weak-field expansion, whether performed in the microscopic or canonical frames, leads to amplitudes depending on the following three quantities: (i) a dimensionless AdS-Planck constant $g^2\equiv (\l \ell_p)^{D-2}$ that can always be taken to obey $g <\!\!\!< 1$ and that counts the order in the perturbative
weak-field expansion, where $\ell_p$ enters via the normalization
of the effective standard action and we are working with
dimensionless physical fields; and (ii) a massive parameter $\l$
that simultaneously (iia) sets the infrared cutoff via $\L\sim
\l^2$ and critical masses $M^2 \sim\l^2$ for the dynamical fields;
and (iib) dresses the derivatives in the interaction vertices thus
enabling the Fradkin-Vasiliev (FV) mechanism;
and {(iii)} the weak-field fluctuation amplitudes\footnote{The gauge-
invariant characterization of the amplitudes is provided by on-shell closed
forms built from $\Phi$ and $A$. A simple set of such ``observables'' are the
zero-form charges found in \cite{Sezgin:2005pv}.}
$|\nabla^n C(\phi)|\sim (\l \ell)^{-s-n} |\phi|$ where $\ell$ is the
characteristic wavelength of the bulk fields.

We stress that what makes higher-spin theory exotic is the dual
purpose served by $\l$ within the FV mechanism whereby positive
and negative powers of $\l$ appear in mass terms and
vertices, respectively. Thus, at each order of the
canonical weak-field expansion scheme, the local bulk interactions -- and in
particular the standard minimal gravitational two-derivative
couplings -- are dominated by strongly coupled top vertices
going like finite positive powers of (energy scale)/(IR cutoff), \emph{i.e.} $(\ell\l)^{-1}$.
On the other hand, in the microscopic weak-field expansion scheme, each order
is given by a potentially divergent Born-Infeld tail, suggesting
that classical solutions as well as amplitudes should be evaluated
directly within the master-field formalism which offers
transparent methods based on requiring associativity of the
operator algebra for setting up and assessing regularized
calculational schemes.

More precisely, the tails are strongly coupled for fluctuations around curved backgrounds that are close to the $AdS_D$ solution, where $\ell\l<\!\!\!< 1$, although it is in principle also possible to expand around backgrounds that are ``far'' from
$AdS_D$, and that might bring in additional new scales altering the nature of
the tails. Sticking to the first background scenario, and remaining with the microscopic frame of fields, Vasiliev's oscillator formalism may offer a natural remedy amounting to
augmenting specific classes of composite operators by associative operator
products. As a result the tails, which are power-series expansions in $z=(\ell\l)^{-1}$ that define special functions in the unphysical region $|z|<\!\!\!< 1$, would be given physically meaningful continuations into the physical region $|z|>\!\!\!> 1$, leading to
microscopic amplitudes ${1\over \ell_p^{D-2}}\int
\widehat V_\L(s_1,\dots,s_N)\sim g^{N-2} \widehat A(s_1,\dots,s_N|\ell\l)$, where $\widehat A(s_1,\dots,s_N|\ell\l)$ are analytically continued amplitudes. If these are bounded uniformly in $\{s_i\}$ for $\ell\l<\!\!\!< 1$, then a semi-classical expansion would be possible if $g<\!\!\!< 1$.

If on the other hand one redefines away the non-minimal tails, and if the higher-derivative nature of the minimal cubic vertices generalizes to $N>3$, then the remaining top vertices ${V}{}^{({\rm top})}_\L(s_1,\dots,s_n)$ would necessarily contain total numbers of derivatives $n_{\rm top}(\{s_i\})$ growing at least linearly with
$\sum_i s_i$, so that the resulting canonical amplitudes $A(s_1,\dots,s_N|\ell\l;g)\sim g^{N-2}(\ell\l)^{-n_{\rm top}(\{s_i\})}$, leaving no room for a uniform semi-classical expansion.

\section{Antifield formulation}
\label{AppBV}
\subsection{Definitions}\label{sec:Def}

In this section we briefly recall the BRST deformation scheme 
\cite{Barnich:1993vg} in the case of spin-$s$ Fronsdal theory, 
that is irreducible and abelian. The containt of the present section 
is mainly based on the works 
\cite{Bekaert:2005ka,Boulanger:2000rq,Bekaert:2005jf}.

According to the general rules of the BRST-antifield formalism, 
a grassmann-odd ghost is introduced, which accompanies each 
grassmann-even gauge parameter of the gauge theory.
It possesses the same algebraic symmetries as the corresponding
gauge parameter. In the cases at hand, it is symmetric and
traceless in its spacetime indices.
Then, to each field and ghost of the spectrum, a corresponding
antifield (or antighost) is added, with the same algebraic symmetries but
the opposite Grassmann parity. 
A $\mathbb{Z}$-grading called {\textit{ghost number}} ($gh$) 
is associated with the BRST differential $s$, while the 
{\textit{antifield number}} ($antigh$) of the antifield $Z^*$ 
associated with the field (or ghost) $Z$ is given by 
$antigh(Z^*)\equiv gh(Z)+1\,$. It is also named
{\textit{ antighost number}}.
More precisely, in the general class of theories under consideration, 
the spectrum of fields (including ghosts)
and antifields together with their respective ghost and antifield 
numbers is given by ($s>2$)
\begin{itemize}
\item the fields 
$\{ A_{\mu}, h_{\mu\nu}, \phi_{\m_1\ldots\m_s} \}\,$ 
with ghost number $0$ and antifield number $0$;
\item the ghosts 
$\{ C, C_{\mu}, C_{\m_1\ldots\m_{s-1}} \}$ 
with ghost number $1$ and antifield number $0$;
\item the antifields $\{ A^{*\mu}, h^{*\mu\nu}, 
\phi^{*\m_1\ldots\m_s} \}$ , 
with ghost number $-1$ and antifield number $1$;
\item the antighosts 
$\{ C^*, C^{*\mu}, C^{*\m_1\ldots\m_{s-1}} \}$  
with ghost number $-2$ and antifield number $2\,$.
\end{itemize}
If the {\textit{pureghost number}} ($pgh$) of an expression simply 
gives the number of ghosts (and derivatives of the ghosts) present
in this expression, the ghost number ($gh$) is simply given by 
$$gh = pgh-antigh\;.$$
The fields and ghosts will sometimes be denoted collectively by $\Phi_I\,$,
the antifields by $\Phi^{*I}$.
\vspace*{.3cm}

The basic object in the antifield formalism is the BRST generator $W_0\,$.
For a spin-$1$ field $A_{\mu}\,$, a spin-$2$ field $h_{\mu\nu}\,$ 
and a (double-traceless) spin-$s$ Fronsdal field 
$\phi_{\mu_1\ldots\mu_s}\,$, it reads
\begin{eqnarray} 
W_{0,1}&=&S_{EM}[A_{\mu}] +\int A^{*\mu}\;\6_\mu C\; d^Dx\;,
\nonumber \\
W_{0,2}&=&S_{PF}[h_{\mu\nu}]+2\,\int h^{*\mu\nu}\;\6_{(\mu}C_{\nu)} \;d^Dx\ ,
\nonumber \\
W_{0,s}&=&S_{F}[\phi_{\mu_1...\mu_s}]
+s\,\int \phi^{*\mu_1...\mu_s}\;\6_{(\mu_1}C_{\mu_2...\mu_s)}\;d^Dx\ .
\nonumber 
\end{eqnarray}
The functional $W_0$ satisfies the master equation $(W_0,W_0)=0$,
where $(\,,)$ is the antibracket given by
\begin{eqnarray}(A,B)=\frac{\d^R A}{\d\Phi_I}
\frac{\d^L A}{\d\Phi^{*I}} - \frac{\d^R A}{\d\Phi^{*I}}
\frac{\d^L A}{\d\Phi_I}\,.
\end{eqnarray}
Let us note that this definition is appropriate for both functionals and 
differentials forms.
In the former case, the summation over $I$ also implies an integration over 
spacetime (de Witt's condensed notation). See the textbook 
\cite{Henneaux:1992ig} for a thorough exposition of the BRST formalism.

The action of the BRST differential $s$ is defined by 
$$sA=(W_0\, , A)\;.$$
The differential $s$ is the sum of the
Koszul-Tate differential $\d$ (which reproduces the equations of motion and 
the Noether identities) and the longitudinal derivative $\g$ (which 
reproduces the gauge transformations and the gauge algebra). Let us write down 
explicitely the action of $\d$ and $\g$ (unless
it is vanishing): For a spin-1 field:
\begin{eqnarray}\d C^* = -\6_\m A^{*\m}\ ,\quad 
\d A^{*\m}=\6_\r F^{\r\m}\ , \quad \g A_\m=\6_\m C\ .\nonumber
\end{eqnarray}
For a spin-2 field:
\begin{eqnarray} \d C^{*\n} = -2\,\6_\m h^{*\m\n}\ , \quad 
\d h^{*\m\n}=-2\, H^{\m\n}\ , \quad \g h_{\m\n}=2\,\6^{}_{(\m} C_{\n)}\ . 
\nonumber
\end{eqnarray}
For a spin-$s$ field:
\begin{eqnarray}
\d C^{*\mu_1...\mu_{s-1}}=-s\left(\6_{\mu_s}\phi^{*\mu_1...\mu_s}-
\frac{(s-1)(s-2)}
{2(D+2s-6)}\;\eta^{(\mu_1\mu_2}\6_{\mu_s}\phi'^{*\mu_3...\mu_{s-1})\mu_s}
\right)\;,
\nonumber
\end{eqnarray}
\begin{eqnarray}\d\phi^{*\mu_1...\mu_s}=G^{\mu_1...\mu_s}\ ,\quad
\g\phi_{\mu_1...\mu_s}=s\,\6_{(\mu_1}C_{\mu_1...\mu_s)}\nonumber\ ,
\end{eqnarray}
where 
$F_{\m\n}=\6_\m A_\n -\6_\n A_\m\,$ and 
$K_{\a\b|\m\n}=-\frac{1}{2}(\6^2_{\a\m}h_{\b\n}+\6^2_{\b\n}h_{\a\m}
-\6^2_{\a\n}h_{\b\m}-\6^2_{\b\m}h_{\a\n})\,$
are the Maxwell field-strength and the linearized Riemann
tensor, respectively\footnote{We use the notation 
$\partial^{N}_{\mu_1\ldots\mu_N}\equiv 
\partial_{\mu_1}\ldots\partial_{\mu_N}\,$.}. 
The linearized Einstein tensor is
$H_{\m\n}=K_{\m\n}-\frac{1}{2}\,\eta_{\m\n}K\,$ where
$K_{\b\n}=\eta^{\a\m}\,K_{\a\b|\m\n}$ is the linearized Ricci 
tensor and $K=\eta^{\b\n}\,K_{\b\n}$ the linearized
scalar curvature. 
Finally, the flat-spacetime spin-$s$ Einstein-like and Fronsdal 
tensors $G_{\mu_1...\mu_s}$ and $F_{\mu_1...\mu_s}$ are
given by 
\begin{eqnarray}
G_{\mu_1...\mu_s}&=&F_{\mu_1...\mu_s}
-\frac{s(s-1)}{4}\,\eta_{(\mu_1\mu_2}\,F'_{\mu_3...\mu_s)}\;,
\nonumber \\
F_{\mu_1...\mu_s}&=&\Box \phi_{\mu_1...\mu_s}
-s\,\6^2_{\r(\mu_1}\phi^{\r}_{\phantom{\r}\mu_2...\mu_s)}+
\frac{s(s-1)}{2}\,\6^2_{(\mu_1\mu_2}\phi'_{\mu_3...\mu_s)}\;.
\nonumber
\end{eqnarray}
For further purposes we also display the spin-$s$ curvature
\begin{eqnarray}
K_{\mu_1\nu_1|...|\mu_s\nu_s}=2^s\;Y^s(\6^s_{\mu_1...\mu_s}
\phi_{\nu_1...\nu_s})\;,\quad s>2\;,
\end{eqnarray}
where we have used the permutation operator
\begin{eqnarray}Y^s=\frac{1}{2^s}\prod_{i=1}^{s} [e-(\mu_i\nu_i)]\;
\nonumber
\end{eqnarray}
that performs total antisymmetrization over the pairs of indices 
$(\mu_i,\nu_i)\,$, $i=1,\ldots,s\,$.
Finally, we note that the Fronsdal and curvature tensors 
are not quite independant. 
The following relations can be established:
\begin{eqnarray}
K^{\rho}_{\phantom{\rho}\nu_{s-1}|\rho\nu_s|\mu_1\nu_1|
\ldots|\mu_{s-2}\nu_{s-2}}
&=& 2^{s-2}\,Y^{s-2}(\6^{s-2}_{\mu_1...\mu_{s-2}}F_{\nu_1...\nu_s})\;,
\nonumber \\
\6^{\mu_s} K_{\mu_1\nu_1|...|\mu_s\nu_s}&=&
2^{s-1}\,Y^{s-1}(\6^{s-1}_{\mu_1...\mu_{s-1}}F_{\nu_1...\nu_s})\;.
\nonumber
\end{eqnarray}

In the following two subsections we 
give some cohomological results needed for the 
BRST-BV analysis of the deformation problem.  

\subsection{Cohomology $H^*(\g)$}\label{Hgamma}

For a proof of general results, see \cite{Bekaert:2005ka}.
The only gauge-invariant functions for a spin-$s$ gauge 
field are functions of 
the field-strength tensor $F_{\m\n}$, the Riemann tensor $K_{\a\b|\m\n}$, 
the Fronsdal tensor $F_{\mu_1...\mu_s}$
and the curvature tensor $K_{\mu_1\nu_1|\mu_2\nu_2|...|\mu_s\nu_s}\,$.
In pureghost number $pgh=0\,$ one has: 
$H^0(\g)=\{f([F_{\mu\nu}],[K],[F_s],[K_s],[\Phi^{*I}])\}$
where the notation $[\psi]$ indicates the (anti)field $\psi$ as
well as all its derivatives up to a finite (but otherwise unspecified)
order. 
In $pgh>0$, it can be shown (along the same lines as in 
\cite{Boulanger:2006gr}, Appendix A)
that one can choose $H^*(\g)$-representatives as the products of 
an element of $H^0(\g)$ with an appropriate number of non $\g$-exact ghosts.
The latter are 
$\{$ $C$, $C_\m$, $\6_{[\m}C_{\n]}$, $C_{\mu_1...\mu_{s-1}}$ $\}$ 
together with the 
traceless part of $Y^j(\6_{\mu_1...\mu_j}C_{\nu_1...\nu_{s-1}})$ for 
$j\leqslant s-1\,$, that we denote 
$U^{(j)}_{\mu_1\nu_1|\ldots\mu_j\nu_j|\nu_{j+1}\ldots\nu_{s-1}}\,$. 
If we denote by $\omega^i_J$ a basis of the products of these 
objects in $pgh=i$,
we get : 
\be H^i(\g)\cong\{\a^J\omega^i_J\ |\ \a^J\in H^0(\g)\}\ .
\ee
More generally, let $\{\o_I\}$ be a basis of the space of polynomials 
in these variables 
(since these variables anticommute, this space is finite-dimensional).
If a local form $a$ is $\gamma$-closed, we have
\begin{eqnarray}
        \g a = 0 \quad\Rightarrow\quad a \,=\,
        \a^J \, \o_J + \g b \,.
\end{eqnarray}
If $a$ has a fixed, finite ghost number, then $a$ can only contain
a finite number of antifields. Moreover, since the
{\textit{local}} form $a$ possesses a finite number of
derivatives, we find that the $\a^J$ are polynomials. Such a
polynomial $\a^J$ will be called  an
{\textit{invariant polynomial}}.\vspace{.3cm}

We shall need several standard results on the cohomology of 
$d$ in the space of invariant polynomials.
\begin{prop}\label{2.2}
In form degree less than $D$ 
and in antifield number strictly greater than $0\,$,
the cohomology of $d$ is trivial in the space of invariant
polynomials.
That is to say, if $\a$ is an invariant polynomial, the equation
$d \a = 0$ with $antigh(\a) > 0$ implies
$ \a = d \b$ where $\b$ is also an invariant polynomial.
\end{prop}
\noindent The latter property is rather generic for gauge theories
(see e.g. Ref. \cite{Boulanger:2000rq} for a proof), as well as
the following:

\begin{prop}\label{csq}
If $a$ has strictly positive antifield number, then the equation
$\gamma a + d b = 0$ is equivalent, up to trivial redefinitions,
to $\gamma a = 0$. More precisely, one can always add $d$-exact
terms to $a$ and get a cocycle $a' := a  + d c$ of $\gamma$, such
that $\g a'= 0$.
\end{prop}

\subsection{Homological groups $H^D_2(\d|d)$ and $H^D_2(\d|d,H^0(\g))$}

We first recall a general result (Theorem 9.1 in \cite{Barnich:1994db}):
\begin{prop}\label{usefll}
For a linear gauge theory of reducibility order $r$,
\begin{eqnarray}
H_p^D(\d \vert\, d)=0\; for\; p>r+2\,. \nonumber
\end{eqnarray}
\end{prop}
\noindent Since the theory at hand has no reducibility, 
we are left with the computation of $H_2^D(\d \vert\, d)\,$.
Then, as we already claimed in \cite{Boulanger:2006gr}, 
for a collection of different spins, 
$H_2^D(\d|d)$ is the direct sum of the homologies of the 
individual cases. 

\noindent For spin$\,$-$1$: 
$$ H_2^D(\d|d)= \left\{\L C^*\,d^Dx \, |\ \L\in\Rn \right\}\,.$$
For spin$\,$-$2$ :
\be H_2^D(\d|d)=\left\{\xi^\m C^{*}_\m \,d^Dx \;
|\ \6_{(\m}\xi_{\n)}=0\right\}\ .
\ee
For spin$\,$-$s$ $(s>2)\,$ (\cite{Bekaert:2005ka,Barnich:2005bn}):
\be H_2^D(\d|d)=\left\{\xi^{\mu_1...\mu_{s-1}}C^*_{\mu_1...\mu_{s-1}}\,d^Dx\ 
|\; \6_{(\mu_1}\xi_{\mu_2...\mu_s)}=0\right\} \;.
\ee


\subsection{BRST deformation}
\label{deformation}
%
As shown in \cite{Barnich:1993vg}, the Noether procedure can be
reformulated within a BRST-cohomological framework. Any
consistent deformation of the gauge theory corresponds to a
solution $$W=W_0+g W_1+g^2W_2+\co(g^3)$$ of the deformed master
equation $(W,W)=0$. Taking into account field-redefinitions, 
the first-order nontrivial
consistent local deformations $W_1=\int a^{D,\,0}$ are in
one-to-one correspondence with elements of the cohomology
$H^{D,\,0}(s \vert\, d)$ of the zeroth order BRST differential
$s=(W_0\,,\cdot)$ modulo the total derivative $d\,$, in maximum
form-degree $D$ and in ghost number $0\,$. That is, one must
compute the general solution of the cocycle condition
\begin{eqnarray}
        s a^{D,\,0} + db^{D-1,1} =0\,,
        \label{coc}
\end{eqnarray}
where $a^{D,\,0}$ is a top-form of ghost number zero and
$b^{D-1,1}$ a $(D-1)$-form of ghost number one, with the
understanding that two solutions of (\ref{coc}) that differ by a
trivial solution should be identified
\begin{eqnarray}
        a^{D,\,0}\sim a^{D,\,0} + s p^{D,-1}  + dq^{D-1,\,0} \nonumber
\end{eqnarray}
as they define the same interactions up to field redefinitions.
The cocycles and coboundaries $a,b,p,q,\ldots\,$ are local forms of
the field variables (including ghosts and antifields).
The corresponding second-order interactions $W_2$ must satisfy the 
consistency condition
$$s  W_2=-\frac{1}{2} (W_1,W_1)\,.$$ 
This condition is controlled by the local BRST cohomology group 
$H^{D,1}(s\vert d)$.
\vspace*{.3cm}

Quite generally, one can expand $a$ according to the antifield
number, as 
\be a=a_0+a_1+a_2+ \ldots a_k\,,
\label{antighdec}\ee
where $a_i$ has antifield number $i$.  The expansion stops at some
finite value of the antifield number by locality, as was proved in
\cite{Barnich:1994mt}.

Let us recall \cite{Henneaux:1997bm} the meaning of the various
components of $a$ in this expansion. The antifield-independent
piece $a_0$ is the deformation of the Lagrangian; $a_1$, which is
linear in the antifields associated with the gauge fields, 
contains the information
about the deformation of the gauge symmetries; 
$a_2$ contains the information
about the deformation of the gauge algebra (the term $C^{*} C C$
gives the deformation of the structure functions appearing in the
commutator of two gauge transformations, while the term $\phi^* \phi^* C
C$ gives the on-shell closure terms); and the $a_k$ ($k>2$) give
the informations about the deformation of the higher order
structure functions and the reducibility conditions.

In fact, using standard reasonings (see e.g. \cite{Boulanger:2000rq}), 
one can remove all components of $a$ with antifield
number greater than 2. 
The key point, as explained e.g. in \cite{Bekaert:2005jf},
is that the invariant 
characteristic cohomology $H^{n,inv}_k(\delta \vert d)$ controls the 
obstructions to
the removal of the term $a_k$ from $a$ and that all 
$H^{n,inv}_k(\delta \vert d)$ vanish for $k>2$ by Proposition 
\ref{usefll} and Theorem \ref{2.6} proved in section \ref{deltamodd4}.  
This proves the first part of the following Theorem \ref{antigh2}, 
valid up to spin $s=4$:
\begin{theorem}
\label{antigh2} Let $a$ be a local top form which is a 
nontrivial solution of the equation (\ref{coc}). 
Without loss of generality, one can assume
that the decomposition (\ref{antighdec}) stops at antighost number
two, i.e. 
\be a=a_0+a_1+a_2\,.
\label{defdecomps}
\ee
Moreover, the element $a_2$ is cubic: linear in the antighosts and quadratic
in the variables 
$\{C,C_{\mu},\partial_{[\mu}C_{\nu]}, C_{\mu_1\ldots\mu_{s-1}},
U^{(j\leqslant s-1)}_{\mu_1\nu_1|\ldots|\mu_j\nu_j|\nu_{j+1}\ldots\nu_{s-1}}
\; \}\vert_{s\leqslant 4}\;$ given in subsection \ref{Hgamma}.  
\end{theorem}
Similarly to (\ref{defdecomps}), one can assume $b=b_0+b_1\,$
in (\ref{coc}) (see e.g. \cite{Boulanger:2000rq}) and insert
the expansions of $a$ and $b$ into the latter equation.  
Decomposing the BRST differential as $s=\d+\g$ yields 
\begin{eqnarray}
\g a_0+\d a_1 +d b_0=0 \,,\label{first}\\
\g a_1+\d a_2 +d b_1=0 \,, \label{second}\\
\g a_2=0\,. \label{minute} 
\end{eqnarray}
The general solution of (\ref{minute}) is given in subsection \ref{Hgamma}. 
\vspace*{.3cm}

\textbf{Remark:} 
Actually, even if the Theorem \ref{2.6} cannot be extended to 
$s>4$ for technical reasons, we can always assume that $a_2$ is
cubic as given in the above Theorem \ref{antigh2}, 
relax the limitation $s\leqslant 4$ 
and proceed with the determination of $a_1$ and $a_0$ according to
(\ref{second}) and (\ref{first}). 
In fact, it is impossible to build a ghost-zero cubic object with
$antigh>2$, so a cubic deformation always stops at $antigh\ 2$. 
Moreover, a cubic element $a_2$ must be proportional to an antighost and quadratic in the ghosts, then, modulo $d$ and $\g$, it is obvious that the only possible cubic deformations are those given in Theorem \ref{antigh2}.
Finally, combining the cohomological approach with other approaches
like the light-cone one \cite{Metsaev:2005ar,Metsaev:2007rn} 
may complete our results, as we actually show in the following. 
Such a combination of two different methods seems to us the most
poweful way to completely solve the first-order deformation problem.

\section{Consistent vertices $V^{\L=0}(1,s,s)$}
\label{sec:1ss}

In this section we use the antifield formalism reviewed above 
and apply it 
to the study of nonabelian interactions between spin-$1$ and spin-$s$ 
gauge fields.
We first examine in detail the interactions of the type $1-2-2\,$, 
and then move on to the general case $1-s-s\,$.

\subsection{Exotic nonabelian vertex $V^{\L=0}(1,2,2)$}
\label{sec:Ftheory}

In this section we show the existence of a cubic cross-interaction between
a spin 1 field and a family of exotic spin 2 fields.
The structure constants of this vertex are antisymmetric,
which is in contradiction with the result for self-interacting spin 2 fields
(see \cite{Boulanger:2000rq}).
In fact, we easily prove that this vertex cannot coexist with the Einstein-
Hilbert theory.

We consider in the following a set of fields in Minkowski spacetime of 
dimension $D\,$.
First, a single electromagnetic field $A_\mu$ with field strength
$F_{\m\n}=\6_\m A_\n -\6_\n A_\m\,$,
invariant under the gauge transformations
$\stackrel{(0)}{\d_\L} A_\m=\6_\m \L\,$.
Then, a family of Pauli-fierz fields $h_{\m\n}^a$ where $a$ is the family 
index.
The linearized Riemann tensor is $K^a_{\a\b|\m\n}=-\frac{1}{2}
(\6^2_{\a\m}h^a_{\b\n}+\6^2_{\b\n}h^a_{\a\m}
-\6^2_{\a\n}h^a_{\b\m}-\6^2_{\b\m}h^a_{\a\n})\,$.
It is invariant under the linearized diffeomorphism gauge transformations
$\stackrel{(0)}{\d_\xi} h^a_{\m\n}=2\6^{}_{(\m}\xi^a_{\n)}\,$.
The linearized Einstein tensor is
$H^a_{\m\n}=K^a_{\m\n}-\frac{1}{2}\eta_{\m\n}K^a\,$, according to the
notation given in Section~\ref{sec:Def}.

The free action is the sum of the electromagnetic action and the different
Pauli-Fierz actions:
\be
S_0 =\int \left( -\frac{1}{4}F_{\m\n}F^{\m\n} - 
h_a^{\m\n}H^a_{\m\n}\right)d^D x\ .
\ee
In order to study the cubic deformation problem efficiently, 
we have used the antifield formalism of \cite{Barnich:1993vg}, 
reviewed in the present paper. 
The antifield formalism allows us to write down
every possible nontrivial deformation of the gauge algebra, 
encoded in the element denoted $a_2$ above.  
It turns out that only one $a_2$ candidate gives rise 
to a consistent vertex $a_0\,$. 
The details of the analysis are relegated to the 
appendix \ref{append122}, not to obscure the reading.
The cubic vertex, gauge transformations and gauge algebra are
\begin{eqnarray}
&\bullet &\; \stackrel{(3)}{\cal{L}} = l_{[ab]}\left[-
F^{\rho\sigma}\6_{[\mu}h^a_{\nu]\r}\6^{\m}h^{b\nu}_{\phantom{b\nu}\s}+2A^{\s}
K^a_{\m\n|\r\s}\6^\m h^{b\n\r}\right]\;,
\nonumber \\
&\bullet &\; \stackrel{(1)}{\d_\xi} A_\rho = 
2l_{[ab]}\6^{}_{[\m}h^a_{\n]\r}\6^{\m}\xi^{b\n}
\nonumber \\
&\bullet &\; \stackrel{(1)}{\d}_{\!\xi,\Lambda}h_{a\nu\rho}= 2l_{[ab]}\left[
\Lambda K^b_{\n\r}+ F^\m_{\phantom{\m}\r} \6_{[\m}\xi^b_{\n]}\right]
\displaystyle -\frac{1}{D-2}l_{[ab]}\eta_{\nu\rho}
\left[\Lambda K^b+F^{\m\n}\6_\m \xi^b_\n\right]
\nonumber \\
&\bullet &\; \left[\stackrel{(0)}{\d}_{\!\xi},\stackrel{(1)}
{\d}_{\!\eta}\right]A_\mu+\left[\stackrel{(1)}{\d}_{\!\xi},\stackrel{(0)}
{\d}_{\!\eta}\right]A_\mu=\6_\mu\Lambda\quad \textrm{where}\quad \Lambda 
=2l_{[ab]}\6_{[\m}\xi^a_{\n]}\6^{\m}\eta^{b\n}
 \;.
\end{eqnarray}

\subsection{Exotic nonabelian vertex $V^{\L=0}(1,s,s)$}

The structure that we have found can be easily extended to obtain a set of 
consistent $1-s-s$ vertices.
Using the notation introduced in Section \ref{sec:Def}, 
the Fronsdal action reads
\begin{eqnarray}
S_{Fs}=\frac{1}{2}\,\int \phi^{\mu_1...\mu_s}_a \,G^a_{\mu_1...\mu_s} 
\,d^Dx\;.
\end{eqnarray}
It is gauge invariant thanks to the Noether identites
$$\6^{\mu_s}G^a_{\mu_1...\mu_s}-\frac{(s-1)(s-2)}
{2(D+2s-6)}\eta_{(\mu_1\mu_2}\6^{\mu_s}
G'^a_{\mu_3...\mu_{s-1})\mu_s}\equiv 0$$
and the symmetry of the second-order differential operator 
defining $G\,$.

The deformation analysis is performed exactly along the same lines as for the 
$1-2-2\,$ vertex.
The uniqueness of the solution has not been proved for spin $s>4$, but we show that it is the only cubic solution deforming the gauge algebra.
The spin-2 solution can ben extended to spin $s$, which leads us to consider a 
deformation of the BRST generator stoping at antighost 2, finishing with the following $a_2$ :
\begin{eqnarray}
a_2=f_{[ab]}C^*Y^{s-1}
(\6^{s-1}_{\mu_1...\mu_{s-1}}C^a_{\nu_1...\nu_{s-1}})
Y^{s-1}(\6^{s-1\mu_1...\mu_{s-1}}
C^{b\nu_1...\nu_{s-1}})d^Dx\,.
\end{eqnarray}
By solving the equation $\d a_2+\g a_1 = d b_1$, we first obtain
\begin{eqnarray}
a_1=\tilde{a}_1+\bar{a}_1=2f_{[ab]}A^{*\r}Y^{s-1}
(\6^{s-1}_{\mu_1...\mu_{s-1}} \phi^a_{\nu_1...\nu_{s-1}\r})Y^{s-1}
(\6^{s-1\mu_1...\mu_{s-1}}C^{b\nu_1...\nu_{s-1}})d^Dx+\bar{a}_1\,
\end{eqnarray}
with $\bar{a}_1\ |\ \g \bar{a}_1=de_1$. 
The resolution of 
$\d a_1+\g a_0=db_0$ provides us with both $\bar{a}_1$ and $a_0\,$:
\begin{eqnarray}
\bar{a}_1&=& 2f_{[ab]}\6^{(s-1)\mu_2...\mu_{s-1}}
\phi^{*a\rho_1\rho_2\nu_3...\nu_{s-1}\t}D^{\nu_1\nu_2\s}_{\rho_1\rho_2\t}
\times\nonumber \\ && \left[F^{\mu_1}_{\phantom{\mu_1}\s} Y^{s-1}
(\6^{s-1}_{\mu_1...\mu_{s-1}}C^b_{\nu_1...\nu_{s-1}})-\frac{1}{2^{s-1}}C 
K^{b\mu_1}_{\phantom{b\mu_1}\s|\mu_1\nu_1|...\mu_{s-1}\nu_{s-1}}
\right]d^Dx
\end{eqnarray}
where
\begin{eqnarray}
D^{\nu_1\nu_2\s}_{\rho_1\rho_2\t}&=&\d^{\nu_1}_{\rho_1}\d^{\nu_2}_{\rho_2}
\d^\s_\t-\frac{1}{2(D+2s-6)}\,\eta_{\rho_1\rho_2}
\eta^{\s\nu_1}\d_\t^{\nu_2}-\frac{s-2}{D+2s-6}\,\eta_{\rho_1\rho_2}
\eta^{\s\nu_2}\d_\t^{\nu_1}\,,
\nonumber \\
a_0 &=& -f_{[ab]}F^{\r\s}Y^{s-1}
(\6^{s-1}_{\mu_1...\mu_{s-1}}\phi^a_{\nu_1...\nu_{s-1}\r})Y^{s-1}
(\6^{{s-1}\mu_1...\mu_{s-1}}
\phi^{b\nu_1...\nu_{s-1}}_{\phantom{b\nu_1...\nu_{s-1}}\s})
d^Dx\nonumber\\&&+f_{[ab]}\frac{1}{2^{s-2}}A^\r
K^a_{\mu_1\nu_1|...|\mu_{s-1}\nu_{s-1}|\r\s}
Y^{s-1}(\6^{{s-1}\mu_1...\mu_{s-1}}
\phi^{b\nu_1...\nu_{s-1}\s})d^Dx\,.
\end{eqnarray}
These components of $W_1$ provide the cubic vertex, the gauge transformations and the gauge algebra:
\begin{eqnarray}
\bullet \stackrel{(3)}{\cl}&=&-f_{[ab]}F^{\r\s}Y^{s-1}(\6^{s-1}_{\mu_1...
\mu_{s-1}}\phi^a_{\nu_1...
\nu_{s-1}\r})Y^{s-1}(\6^{{s-1}\mu_1...\mu_{s-1}}\phi^{b\nu_1...
\nu_{s-1}}_{\phantom{b\nu_1...\nu_{s-1}}\s})\nonumber\\&&+f_{[ab]}
\frac{1}{2^{s-2}}\,A^\r K^a_{\mu_1\nu_1|...|\mu_{s-1}\nu_{s-1}|\r\s}
Y^{s-1}(\6^{{s-1}\mu_1...\mu_{s-1}}\phi^{b\nu_1...\nu_{s-1}\s})\; ,
\end{eqnarray}
\begin{eqnarray}
\bullet\stackrel{(1)}{\d}_\xi A_\mu=Y^{s-1}(\6^{s-1}_{\mu_1...\mu_{s-1}}
\phi^a_{\nu_1...\nu_{s-1}\r})Y^{s-1}(\6^{s-1\mu_1...\mu_{s-1}}
\eta^{b\nu_1...\nu_{s-1}})\end{eqnarray}
\begin{eqnarray}
\bullet\stackrel{(1)}{\d}_{\Lambda,\xi}
\phi_{a\rho_1\rho_2\nu_3...\nu_{s-1}\t}=2(-1)^{s-1}
f_{[ab]}D^{\nu_1\nu_2\s}_{\rho_1\rho_2\t}\times\nonumber 
\6^{s-1\mu_2...\mu_{s-1}}
\Big[ F^{\mu_1}_{\phantom{\mu_1}\s} Y^{s-1}(\6^{s-1}_{\mu_1..\mu_{s-1}}
\xi^b_{\nu_1...\nu_{s-1}})\\ 
 -\frac{1}{2^{s-1}}\Lambda
 K^{b\mu_1}_{\phantom{b\mu_1}\s|\mu_1\nu_1|...|\mu_{s-1}\nu_{s-1}}\Big]
\end{eqnarray}
\begin{eqnarray}
\bullet \left[\stackrel{(0)}{\d}_{\xi},\stackrel{(1)}
{\d}_{\eta}\right]A_\mu+\left[\stackrel{(1)}{\d}_{\xi},\stackrel{(0)}
{\d}_{\eta}\right]A_\mu=\6_\mu \Lambda\end{eqnarray}
where
\begin{eqnarray} \Lambda=2f_{[ab]}Y^{s-1}
(\6^{s-1}_{\mu_1...\mu_{s-1}}\xi^a_{\nu_1...\nu_{s-1}})
Y^{s-1}(\6^{s-1\mu_1...\mu_{s-1}}\eta^{b\nu_1...\nu_{s-1}})\ ,
\end{eqnarray}
the other commutators vanishing.

\subsection{Exhaustive list of interactions $V^{\Lambda=0}(1-s-s)\,$}

The uniqueness of the above cubic nonabelian 
interactions can be obtained by combining the above results 
with those obtained in~\cite{Metsaev:2005ar,Metsaev:2007rn}
using a powerful light-cone method. 
We learn from the work \cite{Metsaev:2005ar}
that there exist only \textit{two} possible
cubic couplings between one spin-$1$ and two spin-$s$ fields. 
The first coupling involves $2s-1$ derivatives in the cubic vertex 
whereas the other involves $2s+1$ derivatives. 
Therefore we conclude that the first coupling corresponds to 
the nonabelian deformation obtained in the previous subsection. 
The other one simply is the Born-Infeld-like coupling
\begin{eqnarray}
 \stackrel{(3)}{\cl}&=&g_{[ab]}\,F^{\r\s}\,
\eta^{\l\t}\,K^a_{\mu_1\nu_1|...|\mu_{s-1}\nu_{s-1}|\r\l}
\;K_{\s\t|}^{b~\;\,\mu_1\nu_1|...|\mu_{s-1}\nu_{s-1}} \; ,
\end{eqnarray}
which is strictly invariant under the abelian gauge transformations.

\section{Uniqueness of the nonabelian $V^{\Lambda=0}(2-4-4)\,$ vertex}
\label{sec:244}

The computation of all the possible nonabelian $2-4-4\,$ or $4-2-2\,$
cubic vertices in Minkowski spacetime of arbitrary dimension
$D>3$ can be achieved along the same lines as for the $1-s-s$ vertex.
We can apply Theorem \ref{antigh2} to find a complete list of the possible $a_2$ terms, thanks to the technical result about $H_k^{n,inv}(\d|d)$ that we provide in Appendix \ref{append244}. Then by solving equations (\ref{second}) and (\ref{first}), we find an unique cubic deformation.

First, it is easily seen that it is impossible to build a non trivial $a_2$ involving one spin $4$ and two spin $2$. Then, in the $2-4-4$ case, the highest number of derivatives allowed for $a_2$ to be nontrivial 
is 6, but Poincar\'e invariance imposes an odd number of derivatives.  Here is 
the only $a_2$ containing 5 derivatives, which gives rise to a consistent 
cubic vertex:
\begin{eqnarray}a_2=f_{AB}C^*_\c U^A_{\a\m|\b\n|\r}
V^{B\a\m|\b\n|\c\r}d^Dx
\nonumber\end{eqnarray} 
where $U^A_{\mu_1\nu_1|\mu_2\nu_2\t}=Y^2(\6^2_{\mu_1\mu_2}
C^A_{\nu_1\nu_2\t})$ and $V^A_{\mu_1\nu_1|\mu_2\nu_2|\mu_3\nu_3}
=Y^3(\6^2_{\mu_1\mu_2\mu_3}C^A_{\nu_1\nu_2\nu_3})$.\\
Then, the inhomogenous solution of $\d a_2+\g a_1=db_1$ can be computed. 
The structure constants have to be symmetric in order for $a_1$ to exist : 
$f_{AB}=f_{(AB)}$
\begin{eqnarray}a_1=\tilde{a}_1+\bar{a}_1=f_{(AB)}\left[
h^{*\phantom{\c}\s}_{\phantom{*}\c}\6^2_{\a\b}\phi^A_{\m\n\r\s}
V^{B\a\m|\b\n|\c\r} - 2 h^{*\g\s} \6^3_{\a\b[\c}\phi^A_{\r]\m\n\s}
U^{B\a\m|\b\n|\r} \right]d^Dx+\bar{a}_1\,.
\nonumber\end{eqnarray}
Finally, the last equation is $\d a_0+\g a_1=db_0$. It allows a solution, 
unique up to redefinitions of the fields and trivial gauge transformations. We 
have to say that the natural writing of $a_2$ and the vertex written in terms 
of the Weyl tensor $w_{\a\b|\c\d}$ do not match automatically. In order to get 
a solution, we first classified the terms of the form $w \6^4(\phi\phi)$. Then 
we classified the possible terms in $\bar{a}_1$, which can be chosen in 
$H^1(\g)$. So they are proportional to the field antifields, proportional to a 
gauge invariant tensor ($K_{PF}$, $F_4$ or $K_4$) and proportional to a non 
exact ghost. Finally, we had to introduce an arbitrary trivial combination in 
order for the expressions to match. The computation cannot be made by hand 
(there are thousands of terms). By using the software FORM \cite{Form}, we 
managed to solve the heavy system of equations and found a consistent set of 
coefficients. We obtained the following $\bar{a}_1$:
\begin{eqnarray}\bar{a}_1=\frac{4}{D+2}\,
f_{AB}\phi^{*A\a}_{\phantom{*A\a}\b}\6^\t K^{\m\n|\a\s} 
U^B_{\m\n|\b\s|\t}d^Dx-2f_{AB}\phi^{*A\m\r}_{\phantom{*A\m\r}\a\b}
\6^\t K^{\a\n|\b\s} U^B_{\m\n|\r\s|\t}d^Dx\nonumber\ .
\end{eqnarray}

and the cubic vertex:
\begin{eqnarray}
a_{0,w}&\approx&f_{AB}w_{\m\n\r\s}\Big[\,
 \frac{1}{2}\,\6^{\m\r\a}\phi'^{A\n\b}\6_\a\phi'^{B\s}_{\phantom{,B\s}\b}
-\frac{1}{3}\,\6^{\m\r\a}\phi^{A\n\b\c\d}\6_\a\phi^{B\s}_{\phantom{B\s}\b\c\d}
+\frac{1}{4}\,\6^{\m\r\a}\phi^{A\n\b\c\d}\6_\b\phi^{B\s}_{\phantom{B\n}\a\c\d}
\nonumber \\ &&
+\frac{3}{4}\,\6^{\m\a\b}\phi'^{A\n\r}\6_\a\phi'^{B\s}_{\phantom{,B\s}\b}
+\frac{3}{4}\,\6^{\m\a\b}\phi^{A\n\r\c\d}\6_\a\phi^{B\s}_{\phantom{B\s}\b\c\d}
-\frac{3}{2}\,\6^{\m\a\b}\phi^{A\n\r}_{\phantom{A\n\r}\b\c}\6_\a\phi'^{B\s\g}
\nonumber \\&&
-\frac{1}{2}\,
\6^{\m}_{\phantom{\m}\b\c}\phi^{A\n\r\a}_{\phantom{A\n\r\a}\d}\6_\a
\phi^{B\s\b\c\d}-\frac{3}{4}\,
\6^{\m\a\b}\phi^{A\s}_{\phantom{A\s}\b\c\d}\6_\a\phi^{B\n\r\c\d}
+\frac{3}{2}\,\6^{\m\a\b}\phi'^{A\s\g}\6_\a\phi^{B\n\r}_{\phantom{B\n\r}\b\c}
\nonumber \\&&
-\,\6^{\m}_{\phantom{\m}\b\c}\phi'^{A\s\a}\6_\a\phi^{B\n\r\b\c}
+\frac{1}{2}\,
\6^{\m}_{\phantom{\m}\b\c}\phi^{A\s\a\c}_{\phantom{A\s\a\c}\d}\6_\a
\phi^{B\n\r\b\d}-\frac{1}{2}\,\6_{\a\b\c}\phi^{A\m\r\a\d}\6_\d\phi^{B\n\s\b\c}
\nonumber \\&&
+\frac{1}{2}\,
\6_{\a\b\c}\phi^{A\m\r\a\t}\6^\b\phi^{B\n\s\c}_{\phantom{B\n\s\c}\t}
+\frac{1}{8}\,\6^{\a\b}\phi'^{A\m\r}\6_{\a\b}\phi'^{B\n\s}
+\frac{3}{8}\,
\6^{\a\b}\phi^{A\m\r\c\d}\6_{\a\b}\phi^{B\n\s}_{\phantom{B\n\s}\c\d}
\nonumber \\&&
-\frac{1}{2}\,\6^{\a}_{\phantom{\a}\b}\phi^{A\m\r\b\c}\6_{\a\c}\phi'^{B\n\s}
+\frac{1}{2}\,
\6_{\a\b}\phi^{A\m\r\b\t}\6^{\a\c}\phi^{B\n\s}_{\phantom{B\n\s}\c\t}
-\frac{3}{4}\,
\6^{\a\b}\phi^{A\m\r\c\d}\6_{\a\c}\phi^{B\n\s}_{\phantom{B\n\s}\b\d}
\nonumber\\&&
+\frac{1}{4}\,\6_{\a\b}\phi^{A\m\r\c\d}\6_{\c\d}\phi^{B\n\s\a\b}\;
\Big]\;,
\label{v244}
\end{eqnarray}
where the weak equality means that we omitted terms that are proportional to
the free field
equations, since they can trivially be absorbed by field redefinitions.
The components $a_1$ and $a_2$ correspond to the following deformation of the gauge 
transformations
\begin{eqnarray}
\stackrel{(1)}{\d}_\xi h_{\s\t}
&=&\frac{1}{2}f_{AB}\Big[\eta_{\t\mu_3}\6^2_{\mu_1\mu_2}
\phi^A_{\nu_1\nu_2\nu_3\s}Y^3(\6^{3\mu_1\mu_2\mu_3}\xi^{B\nu_1\nu_2\nu_3})
\nonumber \\
&& - 2\6^3_{\mu_1\mu_2[\t}\phi^A_{\r]\nu_1\nu_2\s}
Y^2(\6^{2\mu_1\mu_2}\xi^{B\nu_1\nu_2\r})\Big]+ (\s\leftrightarrow\t)
\\
\stackrel{(1)}{\d}_\xi \phi_{\a_1\a_2\a_3\a_4} & = & 
\frac{4}{D+2}\,f_{AB}\eta_{(\a_3\a_4}\d^{\mu_2}_{\a_1}\6^\t
K^{\mu_1\nu_1|\phantom{\a_2}\nu_2}_{\phantom{\mu_1\nu_1|}\a_2)}
Y^2(\6^2_{\mu_1\mu_2}\xi^B_{\nu_1\nu_2\t})
\nonumber \\
&&-2f_{AB}\d^{\mu_1}_{(\a_1}\d^{\mu_2}_{\a_2}\6^\t
K^{\phantom{\a_3}\n_1\phantom{|\a_4)}\n_2}_{\a_3\phantom{\n_1}|\a_4)}
Y^2(\6^2_{\mu_1\mu_2}\xi^B_{\nu_1\nu_2\t})\;.
\end{eqnarray}
and to the following deformation of the gauge algebra:
\begin{eqnarray}\nonumber\left[\stackrel{(0)}{\d}_{\xi},\stackrel{(1)}
{\d}_{\eta}\right]h_{\mu\nu}+\left[\stackrel{(1)}{\d}_{\xi},\stackrel{(0)}
{\d}_{\eta}\right]h_{\mu\nu}=2\6_{(\mu}j_{\nu)}
\end{eqnarray}
where
\begin{eqnarray}j_{\mu_3}=f_{(AB)} 
\6^{2\mu_1\mu_2}\xi^{A\nu_1\nu_2\nu_3}Y^3(\6^3_{\mu_1\mu_2\mu_3}
\eta^B_{\nu_1\nu_2\nu_3}) - (\xi\leftrightarrow\eta)\nonumber
\end{eqnarray}

Let us now consider the other possible cases for $a_2$, containing 3 or 1 
derivatives.
The only possibility with three derivatives is
$a_{2,3}=g_{AB}C^*_\b \6_{[\a}C^A_{\m]\n\r}U^{B\a\m|\b\n|\r}d^Dx$.
Its variation under $\delta$ should be $\g$-closed modulo $d$ but some 
nontrivial terms remain,
of the types $g_{AB}h^*U^A U^B$ and $g_{AB}h^*\6_{[.}C^A_{.]..}V^B$.
The first one can be set to zero by imposing symmetric structure constants,
but the second cannot be eliminated.
The same occurs for one of the candidates with 1 derivative:
$a_{2,1,1}=k_{AB} C^{*\b} C^{A\m\n\r}\6_{[\b} C^B_{\m]\n\r}d^Dx$.
We are then left with 2 candidates involving the spin 4 antifield.
We have found that $\d a_2+\g a_1=d b_1$ can have a solution only
if their structure constants
are proportional :
\begin{eqnarray}
a_{2,1}&=&l_{AB}C^{*A\m\n\r}\left[ C^\a \6_{[\a} C^B_{\m]\n\r}+2\6_{[\m}C_{\a]}C^{B\a}_{\phantom{B\a}\n\r}\right]d^Dx
\nonumber\\
a_{1,1}&=&l_{AB}\phi^{*\m\n\r\s}\left[2h_\s^{\phantom{\s}\a}\6_{[\a} C^B_{\m]\n\r}
-\frac{4}{3}\,C^\a\6_{[\a}\phi^B_{\m]\n\r\s}+8\6_{[\m}h_{\a]\s}C^{B\a}_{\phantom{B\a}\n\r}
-2\6_{[\m} C_{\a]}\phi^{B\a}_{\phantom{B\a}\n\r\s}\right]
\nonumber\\
&& +\frac{2}{D+2}l_{AB}\6_\s\phi'^{*A\r\s}C^\a\phi'^B_{\a\r}\;.\nonumber
\end{eqnarray}
There is no homogenous part $\bar{a}_1$
(because the $\g$-invariant tensors contain at least 2 derivatives).
Then, we have considered the most general expression for $a_0$, which 
is a linear combination of 55 terms of the types $h\phi\6^2\phi$ and 
$h\6\phi\6\phi\,$.
We have found that the equation $\d a_1+\g a_0=db_0$ does not admit any 
solution.
We can conclude that the vertex found with 6 derivatives is the \emph{unique}
nonabelian $2-4-4$ cubic deformation.

This $2-4-4$ vertex, setting $D=4$,
should correspond to the flat limit of the corresponding Fradkin-Vasiliev 
vertex. The uniqueness of the former can be used to prove the uniqueness of 
the latter, as we did explicitly in the $2-3-3$ case.

\section{Consistent vertices $V^{\L=0}(2,s,s)$}
\label{sec:2ss}

\subsection{Nonabelian coupling with $2s-2$ derivatives} 

Our classification of gauge algebra deformations relies on the theorem 
concerning
$H^D_k(\d|d,H^0(\g))$. While apparently obviously true, it actually becomes 
increasingly harder to prove with increasing spin. If
$H^D_k(\d|d,H^0(\g))=H^D_k(\d|d)\cap H^0(\g)$ holds for spin $s>4$ then there 
is only one candidates for a nonabelian type $2-s-s$ deformation, involving $2s-3$ derivatives in $a_2$.

Let us recall that due to the simple expression of $H^D_2(\d|d)$ we are only 
left with a few traceless
building blocks for $a_2$: the antighosts $C^{*\m}$ and 
$C^{*A\mu_1...\mu_{s-1}}$, and a
collection of ghosts and their anti-symmetrized derivatives, namely 
$C_\m$, $\6_{[\m}C_{\n]}$ and tensors 
$U^{(j)A}_{\m_1\n_1|...|\m_j\n_j|\n_{j+1}...\n_{s-1}}$ for $j\leq s-1$, that we have defined in section \ref{Hgamma} \cite{Bekaert:2005ka}. Given this, we can divide the $a_2$ candidates into two categories: those 
proportional to $C^{*A\mu_1...\mu_{s-1}}$ and those proportional to 
$C^{*\mu}$.

The first category is simple to study: $C^{*A\mu_1...\mu_{s-1}}$ carries $s-1$ 
indices, and the
spin 2 ghost can carry at most 2, namely $\6_{[\a}C_{\b]}$. As no traces can be made, the spin 4 ghost can carry at most $s+1$ indices. But
$U^{(2)A}_{\mu_1\nu1|\mu_2\nu_2|\nu_3...\nu_{s-1}}$ contains two antisymmetric 
pairs which
cannot be contracted with $C^{*A}$. The only possible combination involving 
$\6_{[\a}C_{\b]}$
is thus 
$$f_{AB}C^{*A\mu_1...\mu_{s-1}}
\6_{[\mu_1}C_{\a]}C^{B\a}_{\phantom{B\a}\mu_2...\mu_{s-1}}d^Dx\ .$$ 
If we consider the underivated $C_\a$, the only possibility is obviously:
$$g_{AB}C^{*A\mu_1...\mu_{s-1}}C^\a U^{(1)A}_{\a\mu_1|\mu_2...\mu_{s-1}}
d^Dx\ .$$ 
Those two terms contain only one derivative. Just as for the spin 4 case, we 
can show that they are related to an $a_1$ if $f_{AB}=\frac{s}{2}g_{AB}\,$: \begin{eqnarray}a_{2,1}=g_{AB}C^{*A\mu_1...\mu_{s-1}}\left[C^\a \6_{[\a}C^B_{\mu_1]\mu_2...\mu_{s-1}} +\frac{s}{2}\6_{[\mu_1}C_{\a]}C^{B\a}_{\phantom{B\a}\mu_2...\mu_{s-1}}\right]d^n x\ \end{eqnarray} and
\begin{eqnarray}a_{1,1}&=&l_{AB}\phi^{*A\m_1...\mu_s}\left[\frac{s}{2}h_{\m_s}^{\phantom{\m_s}\a}\6_{[\a} C^B_{\m_1]\m_2...\m_{s-1}}-\frac{s}{s-1}C^\a\6_{[\a}\phi^B_{\m_1]\m_2...\m_s}\right.\nonumber\\&&\left.\qquad\qquad\quad\;\;\;+\frac{s^2}{2}\6_{[\m_1}h_{\a]\m_s}C^{B\a}_{\phantom{B\a}\m_2...\m_s}-\frac{s}{2}\6_{[\m_1} C_{\a]}\phi^{B\a}_{\phantom{B\a}\m_2...\m_s}\right]\nonumber\\&&+\frac{s(s-2)}{4(n+2s-6)}l_{AB}\6_\s\phi'^{*A\m_3...\m_s}C^\a\phi'^B_{\a\m_3...\m_{s-1}}\end{eqnarray} Then, the proof of the inconsistency of this candidate is exactly the same as for spin 4. In fact, for every spin $s\geq 4$, there are only 55 possible terms in the vertex. We have thus managed to adapt the proof to spin $s$, this deformation is obstructed.

For the second category, the structure has to be $C^* U^{(i)} U^{(j)}\ ,\ i<j$ 
But $C^*$ carries one index and $U^{(i)}$ carries $i+s-1$. As no traces can be 
taken, it is obvious that $i=j-1$, which leaves us with a family of 
candidates: $$a_{2,2j-1}=l_{AB}C^{*\a}U^{(j-1)A\mu_1\nu_1|...|
\mu_{j-1}\nu_{j-1}|\nu_j...\nu_{s-1}}U^{(j)B}_{\mu_1\nu_1|...|
\mu_{j-1}\nu_{j-1}|\a\nu_j|\nu_{j+1}...\nu_{s-1}}d^Dx$$

Let us now check if these candidates satisfy the equation 
$\d a_2+\g a_1=d b_1$ for some $a_1$. For the second category, we get 
schematically $\d a_2=d(...)+\g(...)+l_{AB}h^* U^{(j)A}U^{(j)B}+l_{AB}h^* 
U^{(j-1)A}U^{(j+1)B}$. The first obstruction can be removed by imposing 
$l_{AB}=l_{(AB)}$ while the second cannot be removed unless $j=s-1$. As the 
tensor $U^{(s)B}$ does not exist, this term is not present at top number of 
derivatives, the second candidate $a_2$ that correspond to an $a_1$ is then:
\begin{eqnarray}
a_{2,2s-3}=l_{AB}C^{*\a}U^{(s-2)A\mu_1\nu_1|...|\mu_{s-2}\nu_{s-2}|
\nu_{s-1}}U^{(s-1)B}_{\mu_1\nu_1|...|\mu_{s-2}\nu_{s-2}|\a\nu_{s-1}}d^Dx
\,. \label{a22ss}
\end{eqnarray}

\subsection{Exhaustive list of cubic $V^{\Lambda=0}(2,s,s)$ couplings}

Using the results of \cite{Metsaev:2005ar}, we learn that there exist
only \textit{three} cubic couplings of the form 
$V^{\Lambda=0}(2,s,s)\,$. They involve
a total number of derivatives in the vertex being respectively 
$2s+2$, $2s$ and $2s-2\,$. Moreover, it is indicated 
\cite{Metsaev:2005ar} that the $\,2s\,$-derivative coupling only 
exists in dimension $D>4\,$. 
{}From our results of the last subsection, 
we conclude that the last coupling is the nonabelian coupling
with $2s-2$ derivatives. The coupling with $2s+2$ derivatives is 
simply the strictly-invariant Born-Infeld-like vertex
\begin{eqnarray}
\stackrel{(3)}{\cl}_{BI}&=& t_{(ab)}\,K^{\a\b|\c\d}\,
\;K_{\a\b|}^{a~\;\,\mu_1\nu_1|...|\mu_{s-1}\nu_{s-1}} 
\;K^b_{\c\d|\mu_1\nu_1|...|\mu_{s-1}\nu_{s-1}} \; ,
\end{eqnarray}
whereas the vertex with $2s$ derivatives is given by 
\begin{eqnarray}
\stackrel{(3)}{\cl}_{2s}&=& u_{(ab)}\;
\delta^{[\mu\n\r\s\l]}_{[\a\b\c\d\e]}\;
h^{\a}_{~\m}
\;K^{a\; \b\c|~~~\!|\mu_1\nu_1|...|\mu_{s-2}\nu_{s-2}}_{~~~\;\;\;\;\n\r}
\;K^{b \;\d\e|}_{~~~\;\;\,\s\l|\mu_1\nu_1|...|\mu_{s-2}\nu_{s-2}}\;.
\end{eqnarray}
It is easy to see that this vertex is not identically zero
and is gauge-invariant under the abelian transformations, up
to a total derivative. 

\section{Summary and Conclusions}

Already for $\L=0$ the notion of minimal coupling needs to be refined to
account for nonabelian vertices with more than two derivatives.
Using the antifield formulation \cite{Barnich:1993vg}, in order to prove that the first nonabelian vertex involving a set $\{\phi^i\}$ of fields is cubic, one needs a technical cohomological result concerning the nature of $H_k(\d|d)$ in the space
of invariant polynomials. This technical result has been obtained
previously up to $s=3$ and has been pushed here up to $s=4\,$
(cfr. Appendix \ref{deltamodd4}).
Supposing that this result holds in the general spin-s case, which 
is equivalent to supposing that the first nonabelian vertex is cubic, 
we have shown in Section \ref{sec:2ss} that there
exist only two possible nonabelian type $2-s-s$ deformations of the gauge
algebra that can be integrated to corresponding gauge transformations.
One of these two candidates has $2s-3$ derivatives and must therefore
give rise to a vertex with $2s-2$ derivatives to be identified with the
flat limit of the corresponding FV $2-s-s$
top vertex~\cite{Fradkin:1986qy,Fradkin:1987ks}.
We have shown that the other candidate is obstructed.
If liftable to a vertex, it would have given the two-derivatives
vertex that corresponds to the minimal Lorentz covariantization.
We have thus proved by cohomological methods what has recently
been obtained by other methods in
\cite{Metsaev:2005ar,Metsaev:2007rn,Porrati:2008rm}.

Then, by combining our cohomological results with those of
Metsaev \cite{Metsaev:2005ar}, we explicitly built the
\emph{exhaustive} list of nontrivial, manifestly covariant vertices
$V^{\Lambda=0}(1,s,s)$ and $V^{\Lambda=0}(2,s,s)\,$,
notifying the relevant information
concerning the nature of the deformed gauge algebra.

For $\L\neq 0$, the standard notion of Lorentz covariantization does apply
although it only provides the bottom vertex of a finite expansion in
derivatives covered by inverse powers of $\L$, whose top vertices therefore
dominate amplitudes (unless extra scales are brought in \emph{e.g.} by
expansions around non-trivial backgrounds). The top-vertices scale with energy
non-uniformly for different spins rendering the standard semi-classical
approach ill-defined unless some additional feature shows up beyond the cubic
level.

Indeed, Vasiliev's fully non-linear higher-spin field equations may provide
such a mechanism whereby infinite tails amenable to re-summation are
developed. The two parallel perturbative expansions in $g$ and $(\ell\l)^{-1}$
resembles those in $g_s$ and $\a'\ell^{-2}$ in string theory, suggesting that
the strong coupling at $(\ell\l)^{-1}>\!\!\!> 1$ corresponds to a tensionless
limit of a microscopic string (or membrane). Indeed, the geometric
underpinning of Vasiliev's equations is that of flat connections and
covariantly constant sections over a base-manifold -- the ``unfold'' - taking
their values in a fiber. This suggests that the total system is described by a
total Lagrangian whose ``pull back'' to the unfold would be a free
differential algebra action (with exterior derivative kinetic terms). On the
other hand, its pull-back to the fiber would be a microscopic quantum theory
in which the Weyl zero-form is subject to master constraints that are
algebraic equations from the unfold point-of-view (thus avoiding the
problematic negative powers of $\ell\l$). A candidate for the microscopic
theory is the tensionless string/membrane in AdS whose phase-space action has
been argued in \cite{Engquist:2005yt,Engquist:2007vj} to be equivalent to a
topological gauged non-compact WZW model with subcritical level
\cite{Engquist:2007pr}.


\vspace{1cm}

{\bf Acknowledgement:} We would like to thank K. Alkalaev, X. Bekaert, C.
Iazeolla, R. Metsaev, M. Porrati, A. Sagnotti, Ph. Spindel and M. Vasiliev for
discussions. The work of N.B. and P.S. is supported in part by the EU
contracts MRTN-
CT-2004-503369 and MRTN-CT-2004-512194 and by the NATO grant PST.CLG.978785.
\vspace*{.3cm}

During the preparation of this manuscript there appeared the work
\cite{Zinoviev:2008ck} that also addresses the issue of the AdS deformation of
the nonabelian 3-3-2 flat-space vertex found in \cite{Boulanger:2006gr}.

\begin{appendix}

\section{The unique $V^{\Lambda=0}(1,2,2)$ vertex}
\label{append122}

\subsection{The gauge algebra, transformations and vertex: 
$a_2$, $a_1$ and $a_0$}

Thanks to he considerations made above, and as Poincar\'e invariance is
required,
the only nontrivial $a_2$ terms are linear in the underivated $antigh\ 2$
antifields and quadratic in the non exact
ghosts. Family indices can be introduced, which allows to multiply the terms
by structure constants. This construction is impossible for 2 spin 1 and 1
spin 2, while there are 3
candidates for 1 spin 1 and 2 spin 2 :
\begin{itemize}
\item $a_{2,1}=f_{[ab]}C^*C_\m^aC^{b\m} d^Dx$
\item $a_{2,2}=g_{ab}C^{*a}_\m C C^{b\m} d^Dx$
\item $a_{2,3}=l_{[ab]}C^*\6_{[\m}C^a_{\n]}\6^{\m}C^{b\n} d^Dx$
\end{itemize}


We must now check if the equation $\d a_2 +\g a_1 = db_1$ admits solutions for
the above candidates.
Let us note that homogenous solutions for $a_1$ have to be considered :
$\g \bar{a}_1=d\bar{b}_1\,$.
Thanks to Proposition \ref{csq}, this equation can be redefined as
$\g \bar{a}_1=0\,$.
The non trival $\bar{a}_1$ are elements of $H(\g)$, and, as they are linear
in the fields, involve at least one derivative.
\begin{eqnarray}
\bullet~\d a_{2,1} &=&-f_{[ab]}\6_\r
A^{*\r}C_\m^aC^{b\m} d^Dx
\nonumber \\
&=& 2 f_{[ab]}A^{*\r}\6_\r C^a_\m C^{b\m}d^Dx+d(...)
\nonumber \\
&=& -\g (f_{[ab]}A^{*\r}h^a_{\r\m}C^{b\m}d^Dx)
+2 f_{[ab]}A^{*\r}\6_{[\r}C^a_{\m]}C^{b\m}d^Dx+d(...)\,.
\end{eqnarray}
The second term can not be $\g$-exact, therefore
the first candidate has to be discarded.
\begin{eqnarray}
\bullet~\d a_{2,2}&=&-2g_{ab}\6_{\n}
h^{*a\m\n}C C^b_\m d^Dx
\nonumber \\
&=& 2g_{ab}h^{*a\m\n}\left[\6_\n C C^b_\m + C \6_\n C^b_\m \right]d^Dx
+ d(...)\nonumber \\
&=& -\g\left(g_{ab}h^{*a\m\n}\left[2A_\m C^b_\n-C h^b_{\m\n}\right]d^Dx
\right) +d(...)\,.
\end{eqnarray}
As there is no homogenous solution with no derivatives,
we can conclude that
\be a_{1,2}=g_{ab}h^{*a\m\n}\left[2A_\m C^b_\n-C h^b_{\m\n}\right]d^Dx
+\g(...)\,.
\ee
Finally, applying the Koszul-Tate differential on the $a_{2,3}$
gives
\begin{eqnarray}
\bullet~\d a_{2,3}&=&-l_{[ab]}\6_\r
A^{*\r}\6_{[\m}C^a_{\n]}\6^{\m}C^{b\n}
d^Dx\nonumber \\
&=& 2 l_{[ab]}A^{*\r}\6^2_{\r[\m}C^a_{\n]}\6^{\m}C^{b\n} d^Dx+d(...)
\nonumber \\
&=& -\g\left(2l_{[ab]}A^{*\r}\6^{}_{[\m}h^a_{\n]\r}\6^{\m}C^{b\n}
d^Dx\right)+d(...)\,.
\end{eqnarray}
Here we may assume the existence of an homogenous solution :
\be
a_{1,3}=\tilde{a}_{1,3}+\bar{a}_{1,3}=2l_{[ab]}A^{*\r}\6^{}_{[\m}
h^a_{\n]\r}\6^{\m}C^{b\n} d^Dx+\bar{a}_{1,3}\ |\ \g \bar{a}_{1,3}=0\,.
\ee


Finally, we can compute the possible vertices $a_0$, that have to be a
solution of $\d a_1+\g a_0=d b_0$ where $a_1$ is one of the above candidates.

For the candidate $a_{1,2}$, we get $\displaystyle\d
a_{1,2}=-2g_{ab}H^{a\m\n}\left[2A_\m C^b_\n-C h^b_{\m\n}\right]d^Dx$.
The second term is
$\g$-exact modulo $d$ if $g_{ab}=g_{[ab]}$ (thanks to the properties of the
Einstein
tensor), but the first one does not work (terms of the form $h^a_{\cdot\cdot}
\6^2_{\cdot\cdot} A_\cdot C_\cdot$ have non vanishing coefficients and are not
$\g$-exact).

Let us now compute the solution $a_{0,3}$, given that $\g
\6_{[\m}h^a_{\n]\r}=\6^2_{\r[\m}C^a_{\n]}$ and
$\6^\r K^a_{\m\n|\r\s}=2\6_{[\m}K^a_{\n]\s}$:
\begin{eqnarray}
\d
\tilde{a}_{1,3}&=&2l_{[ab]}\6_{\s}F^{\s\r}\6_{[\m}h^a_{\n]\r}\6^{\m}C^{b\n}\nonumber\\
&=&2l_{[ab]}\6^{\s}A^{\r}K^a_{\m\n\s\r}\6^{\m}
C^{b\n}+\g\left(l_{[ab]}F^{\s\r}\6_{[\m}h^a_{\n]\r}\6^\m
h^{b\n\s}\right)+d(...)\nonumber\\
&=&-4
l_{[ab]}A^{\r}\6_{[\m}K^a_{\n]\r}\6^{\m}C^{b\n}+4l_{[ab]}C\6_{[\m}
K^a_{\n]\s}\6^{\m}
h^{b\n\s}\nonumber\\&&-\g\left(2l_{[ab]}A^{\r}
K^a_{\m\n|\s\r}\6^{\m}h^{b\n\s}-
l_{[ab]}F^{\s\r}\6_{[\m}h^a_{\n]\r}\6^\m
h^{b\n\s}\right)+d(...)\,.\nonumber
\end{eqnarray}
The first two terms are $\delta$-exact and correspond to a nontrivial
$\bar{a}_{1,3}$.
The last two terms are the vertex:
\begin{eqnarray}
a_{0,3}&=&l_{[ab]}\left[-
F^{\rho\sigma}\6_{[\mu}h^a_{\nu]\r}\6^{\m}h^{b\nu}_{\phantom{b\nu}\s}
+2A^{\s} K^a_{\m\n|\r\s}\6^\m h^{b\n\r}\right]d^Dx\,,
\nonumber \\
\bar{a}_{1,3}&=&2l_{[ab]}h^{*a\n\r}\left[CK^b_{\n\r}+
F^\m_{\phantom{\m}\r} \6_{[\m}C^b_{\n]}\right]-\frac{1}{D-2}l_{[ab]}
h^{*a}\,\!'\left[CK^b+F^{\m\n}\6_\m C^b_\n\right]+\g(...)\,.
\end{eqnarray}

\subsection{Inconsistency with Einstein-Hilbert theory}

Here we show that, as expected, the spin-2 massless
fields considered in the previous section cannot be considered as
the linearized Einstein-Hilbert graviton.
Let us consider the second order in $g$ in the master equation : it can be
written
$(W_1,W_1)=-2sW_2$. Let us decompose $W_2$ according to the antighost number :
$W_2=\int(c_0+c_1+c_2+...)$. We will just check here the highest antighost
part :
$(a_2,a_2)=-2\g c_2-2\d c_3+d(...)$. But indeed, in any theory in which $a_2$
is linear in
the antighost 2 antifields and quadratic in the ghosts, $(a_2,a_2)$ cannot
depend on the
antighost 1 antifields or on the fields, so that no $\d c_3$ can appear. This
indicates
that the expansion of $W_2$ stops at antighost 2 for those theories.
But in fact, we get here :
$$(a_{2,3},a_{2,3})=2\frac{\d a_{2,3}}{\d C^{*\mu}_a}
\frac{\d a_{2,3}}{\d C^a_\mu}+2\frac{\d a_{2,3}}{\d C^*}
\frac{\d a_{2,3}}{\d C}=0\quad .$$
This just means that the solution that we found is self-consistent at
that order.
But we also have to check the compatibility with self-interacting spin 2
fields. Let us
consider the $a_2$ for a collection of Einstein-Hilbert theories (this can be
found in \cite{Boulanger:2000rq}) :
$a_{2,EH}=f_{(abc)}C^{*a\mu}C^{b\nu}\6_{[\mu}C^c_{\nu]}$, in
which the coefficients $f_{abc}$ can be chosen diagonal. Let us now compute
\begin{eqnarray}(a_{2,EH},a_{2,3})&=&\frac{\d a_{2,EH}}{\d C^{*\r}_e}
\frac{\d a_{2,3}}{\d
C^e_\r}=-2f^e_{bc}l_{ea}C^{b\n}\6_{[\r}C^c_{\n]}\6_\t\left[C^*\6^{[\t}C^{|a|
\r]}\right]\nonumber\\&=&\g(...)+d(...)-2f^e_{bc}l_{ae}C^*\eta^{\s\n}\6_{[\t}
C^b_{\s]}\6_{[\r}C^c_{\n]}\6^{[\t}C^{|a|\r]}\,.
\end{eqnarray}
This can be consistent only if $f^e_{(bc}l_{a)e}=0$. But if we choose
$f_{abc}$ diagonal,
we obtain $f^a_{aa}l_{ba}=-2f^a_{ab}l_{aa}=0$, which means that the $f$'s or
the $l$'s have to vanish. In other words, the spin 2 particles interacting
with the spin 1 in our vertex cannot be Einstein-Hilbert gravitons.

\section{The unique $V^{\Lambda=0}(2,4,4)$ nonabelian vertex}
\label{append244}

\subsection{Invariant cohomology of $\d$ modulo $d$ for spin 4}
\label{deltamodd4}

The following theorem is crucial, in the sense that it enables one
to prove the uniqueness of the deformations,
within the cohomological approach of \cite{Barnich:1993vg}:
\begin{theorem}\label{2.6}
Assume that the invariant polynomial $a_{k}^{p}$
($p =$ form-degree, $k =$ antifield number) is $\delta$-trivial modulo $d$,
\begin{eqnarray}
a_{k}^{p} = \delta \mu_{k+1}^{p} + d \mu_{k}^{p-1} ~ ~ (k \geqslant 2).
\label{2.37}
\end{eqnarray}
Then, one can always choose $\mu_{k+1}^{p}$ and $\mu_{k}^{p-1}$ to be
invariant.
\end{theorem}
To prove the theorem, we need the following lemma, proved
in~\cite{Boulanger:2000rq}.
\begin{lemma} \label{l2.1}
If $a $ is an invariant polynomial that is $\delta$-exact, $a = \d b$,
then, $a $ is $\delta$-exact in the space of invariant polynomials.
That is, one can take $b$ to be also invariant.
\end{lemma}
The proof of Theorem \ref{2.6} for spin-4 gauge field proceeds in
essentially the same way as for the spin-3 case presented in detail in
\cite{Bekaert:2005jf}, to which we refer for the general lines
of reasoning. We only give here the piece of proof where things
differ significantly from the spin-3 case.
\vspace*{.8cm}

Different situations are considered,
depending on the values of $p$ and $k$.
In form degree $p<D\,$, the proof goes as in \cite{Bekaert:2005jf}.
In form degree $p=D\,$, two cases must be considered:
$k>D$ and $k\leqslant D\,$.
In the first case, the proof goes as in \cite{Bekaert:2005jf},
the new features appearing when $p=D\,$ and $k\leqslant D\,$.
Rewriting the top equation (i.e. (\ref{2.37}) with $p=D$) in dual
notation, we have
\be a_k=\d b_{k+1}+\pa_{\r}j^{\r}_{k},~ (k\geqslant 2).
\label{2.44} \ee
We will work by induction on the antifield
number, showing that if the property expressed in Theorem \ref{2.6} is true
for $k+1$ (with $k>1$),
then it is true for $k$. As we already know that it is
true in the case $k>D$, the theorem will be proved.\vspace{.4cm}

\noindent{\bf{Inductive proof for $k\leqslant D$}} : The proof follows the
lines of
Ref. \cite{Barnich:1994mt} and decomposes in two parts. First, all
Euler-Lagrange derivatives of (\ref{2.44}) are computed. Second, the Euler-
Lagrange (E.L.) derivative of an invariant quantity is also invariant. This
property is used to express the E.L. derivatives of $a_k$ in terms of
invariants only. Third, the homotopy formula is used to reconstruct $a_k$ from
his E.L. derivatives. This almost ends the proof.\vspace{1.5mm}

{\bf (i)} Let us take the E.L. derivatives of (\ref{2.44}). Since the E.L.
derivatives with respect to $C^*_{\a\b\c}$,
the antifield associated with the ghost $C^{\a\b\c}$, commute with $\d$,
we get first :
\begin{eqnarray}
\frac{\d^L a_k}{\d C^*_{\a\b\c}} =\d Z^{\a\b\c}_{k-1}
\label{2.45}
\end{eqnarray}
with
$Z^{\a\b\c}_{k-1}=\frac{\d^L b_{k+1}}{\d C^*_{\a\b\c}}$.
For the E.L. derivatives of $b_{k+1}$ with respect to $h^*_{\m\n\r\s}$
we obtain, after a direct computation,
\begin{eqnarray}
\frac{\d^L a_k}{\d h^*_{\m\n\r\s}}=-\d X^{\m\n\r\s}_k +
4\pa^{(\m}Z^{\n\r\s)}_{k-1}.
\label{2.46}
\end{eqnarray}
where $X^{\m\n\r\s}_{k}=\frac{\d^L b_{k+1}}{\d h^*_{\m\n\r\s} }$.
Finally, let us compute the E.L. derivatives of $a_k$ with respect to
the fields.
We get :
\begin{eqnarray}
\frac{\d^L a_k}{\d h_{\m\n\r\s}}=\d Y^{\m\n\r\s}_{k+1} +
{\cg}^{\m\n\r\s\vert\a\b\g\d}
X_{\a\b\g\d\vert k}
\label{2.47}
\end{eqnarray}
where $Y^{\m\n\r\s}_{k+1}=\frac{\d^L b_{k+1}}{\d h_{\m\n\r\s}}$ and
${\cg}^{\m\n\r\s\vert\a\b\g\d}(\partial)$ is the second-order self-adjoint
differential operator appearing in Fronsdal's equations of motion
$0=\frac{\d S^F[h]}{\d h_{\m\n\r\s}}\equiv
G^{\m\n\r\s}={\cg}^{\m\n\r\s\vert\a\b\g\d}\,h_{\a\b\g\d}\,.$
The hermiticity of $\cg$ implies
${\cg}^{\m\n\r\s\vert\a\b\g\d}={\cg}^{\a\b\g\d\vert\m\n\r\s}$.\vspace{2mm}

{\bf (ii)} The E.L. derivatives of an invariant object are invariant. Thus,
$\frac{\d^L a_k}{\d C^*_{\a\b\c}}$ is invariant.
Therefore, by Lemma \ref{l2.1} and Eq. (\ref{2.45}), we have also
\begin{eqnarray}
\frac{\d^L a_k}{\d C^*_{\a\b\c}} =\d Z'^{\a\b\c}_{k-1}
\label{2.45'}
\end{eqnarray}
for some invariant $Z'^{\a\b\c}_{k-1}$.
Indeed, let us write the decomposition
$Z^{\a\b\c}_{k-1} = Z'^{\a\b\c}_{k-1} + {\tilde{Z}}^{\a\b\c}_{k-1}$,
where ${\tilde{Z}}^{\a\b\c}_{k-1}$ is obtained from
${Z}^{\a\b\c}_{k-1}$ by setting to zero all
the terms that belong only to $H(\g)$.
The latter operation clearly commutes with taking the $\d$ of something,
so that Eq. (\ref{2.45}) gives $0 = \d {\tilde{Z}}^{\a\b\c}_{k-1}$ which,
by the acyclicity of $\d$, yields
${\tilde{Z}}^{\a\b\c}_{k-1}=\d \s_k^{\a\b\c}$ where
$\s_k^{\a\b\c}$ can be chosen to be traceless.
Substituting $\d \s_k^{\a\b\c} + Z'^{\a\b\c}_{k-1}$ for
$Z^{\a\b\c}_{k-1}$ in Eq. (\ref{2.45}) gives Eq. (\ref{2.45'}).

Similarly, one easily verifies that
\begin{eqnarray}
\frac{\d^L a_k}{\d h^*_{\m\n\r\s}}=-\d X'^{\m\n\r\s}_k +
4\pa^{(\m}Z'^{\n\r\s)}_{k-1}\,,
\label{2.46'}
\end{eqnarray}
where $X^{\m\n\r\s}_k = X'^{\m\n\r\s}_k + 4\pa^{(\m}\s^{\n\r\s)}_{k} +
\d \r_{k+1}^{\m\n\r\s}$.
Finally, using ${\cg}^{\m\n\r\s}{}_{\a\b\g\d}\,\pa^{(\a}\s^{\b\g\d)}{}_k=0$
due to the gauge invariance of the equations of motion
($\s_{\a\b\d}$ has been taken traceless), we find
\begin{eqnarray}
\frac{\d^L a_k}{\d h_{\m\n\r\s}} = \d Y'^{\m\n\r\s}_{k+1}
+{\cg}^{\m\n\r\s}{}_{\a\b\g\d}{X'}^{\a\b\g\d}_k
\label{2.47'}
\end{eqnarray}
for the invariants $X'^{\m\n\r\s}_k$ and $Y'^{\m\n\r\s}_{k+1}$.
Before ending the argument by making use of the homotopy formula,
it is necessary to know more about the invariant $Y'^{\m\n\r\s}_{k+1}$.

Since $a_k$ is invariant, it depends on the fields only
through the curvature $K$, the Fronsdal tensor and their derivatives.
(We substitute $4\,\pa^{[\d}\pa_{[\g}F^{~\,\s]}_{\r]~~\m\n}$ for
$\eta^{\a\b}K^{\d\s}_{~~\,|\a\m|\b\n|\g\r}$ everywhere.)
We then express the Fronsdal tensor in terms of the
Einstein tensor:
$F_{\m\n\r\s} = G_{\m\n\r\s} - \frac{6}{n+2}\,\eta_{(\m\n}G_{\r\s)}$,
so that we can write $a_k = a_k([\Phi^{*i}],[K],[G])$\,, where $[G]$
denotes the Einstein tensor and its derivatives.
We can thus write
\begin{eqnarray}
\frac{\d^L a_k}{\d h_{\m\n\r\s}} = {\cg}^{\m\n\r\s}{}_{\a\b\g\d}
{A'}^{\a\b\g\d}_k + \pa_{\a}\pa_{\b}\pa_{\g}\pa_{\d}
{M'}^{\a\m\vert\b\n\vert\g\r\vert\d\s}_k
\label{2.49}
\end{eqnarray}
where $${A'}^{\a\b\g\d}_k\propto\frac{\d a_k}{\d G_{\a\b\g\d}}$$ and
$${M'}_k^{\a\m\vert\b\n\vert\g\r\vert\d\s}\propto{\frac{\d a_k}
{\d K_{\a\m\vert\b\n\vert\g\r\vert\d\s}}}$$
are both invariant and respectively have the same symmetry properties as
the ``Einstein" and ``Riemann" tensors.

Combining Eq. (\ref{2.47'}) with Eq. (\ref{2.49}) gives
\begin{eqnarray}
\d Y'^{\m\n\r\s}_{k+1} =
\pa_\a\pa_\b\pa_\g\pa_\d{M'}_k^{\a\m\vert\b\n\vert\g\r\vert\d\s}
                       + {\cg}^{\m\n\r\s}{}_{\a\b\g\d} {B'}^{\a\b\g\d}_k
\label{2.50}
\end{eqnarray}
with ${B'}^{\a\b\g\d}_k:={A'}^{\a\b\g\d}_k-{X'}^{\a\b\g\d}_k$.
Now, only the first term on the right-hand-side of Eq. (\ref{2.50}) is
divergence-free,
$\pa_{\m}(\pa_{\a\b\g}{M'}_k^{\a\m\vert\b\n\vert\g\r})\equiv 0$,
not the second one which instead obeys a relation analogous to the Noether
identities
\begin{eqnarray}
\partial^{\tau}G_{\m\n\r\tau}-\frac{3}{(n+2)}\,
\eta_{(\m\n}\partial^{\tau}G'_{\r)\tau}=0\,.
\nonumber
\end{eqnarray}
As a result, we have
$\d\Big[\pa_{\m}({Y'}^{\m\n\r\s}_{k+1}-\frac{3}{D+2}\,\eta^{(\n\r}
{Y'}_{k+1}^{\s)\m})\Big]=0\,$,
where ${Y'}_{k+1}^{\m\s}\equiv \eta_{\n\r}{Y'}_{k+1}^{\m\n\r\s}\,$.
By Lemma~\ref{l2.1}, we deduce
\begin{eqnarray}
        \pa_{\m}\Big({Y'}^{\m\n\r\s}_{k+1}-\frac{3}{D+2}\,\eta^{(\n\r}
{Y'}_{k+1}^{\s)\m}\Big)
        +\d {F'}_{k+2}^{\n\r\s}=0
        \,, \label{truc}
\end{eqnarray}
where ${F'}_{k+2}^{\n\r\s}$ is invariant and can be chosen symmetric and
traceless.
Eq. (\ref{truc}) determines a cocycle of $H^{D-1}_{k+1}(d\vert\d)$, for given
 $\n$, $\r$ and $\s\,$. Using the general isomorphisms
 $H^{D-1}_{k+1}(d\vert\d)\cong H^{D}_{k+2}(\d\vert d)\cong 0$ ($k\geqslant 1$)
 \cite{Barnich:1994db} we deduce
\begin{eqnarray}
        {Y'}^{\m\n\r\s}_{k+1}-\frac{3}{D+2}\,\eta^{(\n\r}{Y'}_{k+1}^{\s)\m}=
        \pa_{\a}T_{k+1}^{\a\m\vert\n\r\s} + \delta P^{\m\n\r\s}_{k+2}
        \,, \label{truc2}
\end{eqnarray}
where both $T^{\a\m\vert\n\r\s}_{k+1}$ and $P^{\m\n\r\s}_{k+2}$ are
invariant by the induction hypothesis. Moreover,
$T^{\a\m\vert\n\r\s}_{k+1}$ is antisymmetric in its first two
indices. The tensors $T^{\a\m\vert\n\r\s}_{k+1}$ and
$P^{\m\n\r\s}_{k+2}$ are both symmetric and traceless in $(\n,\r,\s)$.
This results easily from taking the trace of Eq. (\ref{truc2}) with
$\eta_{\n\r}$ and using the general isomorphisms
$H^{D-2}_{k+1}(d\vert\d)\cong H^{D-1}_{k+2}(\d\vert d)\cong
H^{D}_{k+3}(\d\vert d)\cong 0$ \cite{Barnich:1994db} which hold
since $k$ is positive. From Eq. (\ref{truc2}) we obtain
\begin{eqnarray}
        {Y'}^{\m\n\r\s}_{k+1} =
        \pa_{\a} [ T_{k+1}^{\a\m\vert\n\r\s}+
\frac{3}{D}\,T_{k+1}^{\a\vert\m(\n}\eta^{\r\s)} ]
        + \delta (...)
         \label{truc3}
\end{eqnarray}
where $T_{k+1}^{\a\vert\m\n}\equiv \eta_{\t\r}T_{k+1}^{\a\t\vert\r\m\n}\,$.
We do not explicit the $\delta$-exact term since it plays no role in the
following.
Since $Y'^{\m\n\r\s}_{k+1}$ is symmetric in $\m$ and $\n$, we have also
$$\pa_{\a}\Big(T_{k+1~~\r\s}^{\a[\m\vert\n]}+
\frac{2}{D}\,T_{k+1~(\s}^{\a\vert[\m}\d^{\n]}_{\r)}\Big)
+\;\delta (...)=0\,.$$
The triviality of $H^{D}_{k+2}(d \vert \d)$ ($k>0$) implies again that
$T_{k+1~~\r\s}^{\a[\m\vert\n]}+\frac{2}
{D}\,T_{k+1~(\s}^{\a\vert[\m}\d^{\n]}_{\r)}$
is trivial, in particular,
\begin{eqnarray}
\partial_{\beta}S^{'\beta\a|\mu\nu|}_{\qquad~\r\s} + \d(...) =
T_{k+1~~\r\s}^{\a[\m\vert\n]}+\frac{2}{D}\,
T_{k+1~(\s}^{\a\vert[\m}\d^{\n]}_{\r)}
\label{derStoT}
\end{eqnarray}
where $S^{'\beta\a|\mu\nu|}_{\qquad~\r\s}$
is antisymmetric in the pairs of indices ($ \b, \a$) and ($\m,\n$),
while it is symmetric and traceless in ($ \r, \s$).
Actually, it is traceless in $\m, \n, \r\,\s$ as the right-hand side of
the above equation shows.
The induction assumption allows us to choose
$S^{'\beta\a|\mu\nu|}_{\qquad~\r\s}$, as well as the quantity under
the Koszul-Tate differential $\d\,$.
We now project both sides of Eq. (\ref{derStoT}) on
the following irreducible representation of the orthogonal group
\begin{picture}(35,16)(0,0)
\multiframe(1,4)(10.5,0){3}(10,10){$\a$}{$\m$}{$\s$}
\multiframe(1,-6.5)(10.5,0){2}(10,10){$\r$}{$\n$}
\end{picture}
and obtain
\begin{eqnarray}
        \partial_{\b}W^{'\b|\a\r|\m\n|\s}_{k+1}+ \d (\dots) = 0
\label{derW}
\end{eqnarray}
where $W^{'\b|\a\r|\m\n|\s}_{k+1}$ denotes the corresponding projection
of $S^{'\beta\a|\mu\nu|\r\s}\,$.
Eq. (\ref{derW}) determines, for given $(\m, \n, \a, \r,\s)\,$,
a cocycle of $H^{D-1}_{k+1}(d\vert\d,H(\gamma))$.
Using again the isomorphisms \cite{Barnich:1994db}
$H^{D-1}_{k+1}(d\vert\d)\cong H^{D}_{k+2}(\d\vert d)\cong 0$ ($k\geqslant 1$)
and the induction hypothesis, we find
\begin{eqnarray}
W^{'\b|\a\r|\m\n|\s}_{k+1}
 = \pa_{\l}\phi^{\l\b\vert\a\r|\m\n|\s}_{k+1} +
        \d (\dots)
        \label{Wintermsofphi}
\end{eqnarray}
where $\phi^{\l\b\vert\a\r|\m\n|\s}_{k+1}$ is invariant,
antisymmetric in $(\l, \b)$ and possesses the irreducible,
totally traceless symmetry
\begin{picture}(35,16)(0,0)
\multiframe(1,4)(10.5,0){3}(10,10){$\a$}{$\m$}{$\s$}
\multiframe(1,-6.5)(10.5,0){2}(10,10){$\r$}{$\n$}
\end{picture}
in its last five indices. The $\d$-exact term is invariant as well.
Then, projecting the equation (\ref{Wintermsofphi}) on the
totally traceless irreducible representation
\begin{picture}(35,16)(0,0)
\multiframe(1,4)(10.5,0){3}(10,10){$\a$}{$\m$}{$\s$}
\multiframe(1,-6.5)(10.5,0){3}(10,10){$\r$}{$\n$}{$\b$}
\end{picture}
and taking into account that $W^{'\b|\a\r|\m\n|\s}_{k+1}$
is built out from $S^{'\beta\a|\mu\nu|\r\s}\,$, we find
\begin{eqnarray}
\partial_{\l}\Psi^{'\l|\a\r|\m\n|\s\b}_{k+1} + \d (\dots)
 = 0
\end{eqnarray}
where $\Psi^{'\l|\a\r|\m\n|\s\b}_{k+1}$ denotes the
corresponding projection of $\phi^{\l\b\vert\a\r|\m\n|\s}_{k+1}\,$.
The same arguments used before imply
\begin{eqnarray}
\Psi^{'\l|\a\r|\m\n|\s\b}_{k+1} = \partial_{\t}
\Xi^{'\t\l|\a\r|\m\n|\s\b} + \delta(...)
\label{PsitoXi}
\end{eqnarray}
where the symmetries of $\Xi^{'\t\l|\a\r|\m\n|\s\b}$ on its
last 6 indices can be read off from the left-hand side and
where the first pair of indices is antisymmetric.
Again, $\Xi^{'\t\l|\a\r|\m\n|\s\b}$ can be taken to be invariant.

Then, we take the projection of $\Xi^{'\t\l|\a\r|\m\n|\s\b}$ on
the irreducible representation
\begin{picture}(45,16)(0,0)
\multiframe(1,4)(10.5,0){4}(10,10){$\t$}{$\a$}{$\m$}{$\s$}
\multiframe(1,-6.5)(10.5,0){4}(10,10){$\l$}{$\r$}{$\n$}{$\b$}
\end{picture}
of $GL(D)$ (here we do not impose tracelessness) and denote the
result by $\Theta^{'\t\l|\a\r|\m\n|\s\b}\,$. This invariant tensor possesses
the algebraic symmetries of the invariant spin-$4$ curvature tensor.
Finally, putting all the previous results together,
we obtain the following relation, using the symbolic
manipulation program \emph{Ricci} \cite{Lee}:
\be
        6\,{Y'}^{\m\n\r\s}_{k+1} =
\pa_{\a}\pa_{\b}\pa_{\c}\pa_{\d}\Theta^{'\a\m\vert\b\n\vert\c\r|\d\s}_{k+1}
               + {\cg}^{\m\n\r\s}{}_{\a\b\g\d}
  \widehat{X}_{k+1}^{'\a\b\g\d}{}+\d(\ldots)\,,
\label{result1}
\ee
with
\begin{eqnarray}
        \widehat{X}^{'}_{\a\b\g\d\vert k+1} &:=&
\frac{\cy^{\m\n\r\s}_{\a\b\c\d}}{D-2}\,
    \Big[-\frac{1}{3}\,\eta^{\t\l}
   S_{\t\m|\l\n|\r\s|k+1} + \frac{1}{3(D+1)}\,
\eta_{\m\n}\eta^{\t\l}\eta^{\kappa\z}
 (S_{\t\kappa|\l\z|\r\s|k+1} + 2\,S_{\t\kappa|\l\r|\z\s|k+1})
\nonumber \\
&&+\;\frac{2(D-2)}{D}\,\eta^{\kappa\t}\partial^{\l}
\phi_{\kappa\m|\l\n|\t\r|\s}
-\frac{4(D-2)}{D(D+2)}\,\eta_{\m\n}\eta^{\kappa\t}\eta^{\x\z}\partial^{\l}
\phi_{\kappa\x|\t\z|\l\m|\n}\Big]\;
        \label{result2}
\end{eqnarray}
being double-traceless and where $\cy^{\m\n\r\s}_{\a\b\c\d}$
projects on completely symmetric rank-4 tensors.\vspace{2mm}

{\bf (iii)} We can now complete the argument. The homotopy formula
\begin{eqnarray}
a_k = \int^{1}_{0}dt\,\left[C^*_{\a\b\c}\frac{\d^L a_k}{\d C^*_{\a\b\c}}+
h^*_{\m\n\r\s}\frac{\d^L a_k}{\d h^*_{\m\n\r\s}}+
h^{\m\n\r\s}\frac{\d^L a_k}{\d h^{\m\n\r\s}}\right](th\,,\,th^*\,,\,tC^*)
\label{homotopy}
\end{eqnarray}
enables one to reconstruct $a_k$ from its Euler-Lagrange derivatives.
Inserting the expressions (\ref{2.45'})-(\ref{2.47'}) for
these E.L. derivatives, we get
\begin{eqnarray}
a_k=\d\Big(\int^{1}_{0}dt\,[C^*_{\a\b\c}Z'^{\a\b\c}_{k-1}
+h^*_{\m\n\r\s}X'^{\m\n\r\s}_{k}+h_{\m\n\r\s}Y'^{\m\n\r\s}_{k+1}](t)\,\Big)
+\pa_{\r}k^{\r}.\label{invhf}
\end{eqnarray}
The first two terms in the argument of $\d$ are manifestly invariant.
In order to prove that the third term can be assumed to be invariant
in Eq. (\ref{invhf}), we use Eq. (\ref{result1}) to find that (absorbing the
irrelevant factor 6 in a redefintion of $Y'^{\m\n\r\s}$)
$$h_{\m\n\r\s}\,Y'^{\m\n\r\s}_{k+1}=\frac{1}{16}\,
\Theta^{'\a\m\vert\b\n\vert\c\r\vert\d\s}_{k+1}
K_{\a\m\vert\b\n\vert\c\r\vert\d\s}
+G_{\a\b\g\d}\widehat{X}^{'\a\b\g\d}{}_{k+1}+\pa_\r \ell^\r+\d(\ldots)\,,$$
where we integrated by part four times in order to get the first term of the
r.h.s. while
the hermiticity of ${\cg}^{\m\n\r\s\vert\a\b\g\d}$ was used to obtain the
second term.

We are left with $a_k = \d \m_{k+1} +
\pa_{\r}\n^{\r}_k\,$, where $\m_{k+1}$ is invariant. That
$\n^{\r}_{k}$ can now be chosen invariant is straightforward.
Acting with $\g$ on the last equation yields $\pa_{\r} (\g
{\n}^{\r}_{k}) =0\,$. By the Poincar\'e lemma, $\g {\n}^{\r}_{k} =
\pa_{\s} (\t_k^{[\r \s]})\,$. Furthermore, Proposition \ref{csq}
concerning $H(\g\vert\, d)$ at positive antighost number $k$ implies that
one can redefine $\n^{\r}_{k}$ by the addition of trivial
$d$-exact terms such that one can assume $\g {\n}^{\r}_{k}=0\,$.
As the pureghost number of ${\n}^{\r}_{k}$ vanishes, the last
equation implies that $\n^{\r}_{k}$ is an invariant
polynomial.

\subsection{Uniqueness of the $V^{\Lambda=0}(2,4,4)\,$ vertex}

As the main theorems are now established up to spin 4, we can classify the
nontrivial $a_2$ terms, which correspond to the deformations of the gauge
algebra. The highest number of derivaties allowed for $a_2$ to be nontrivial
is 6, but Poincar\'e invariance imposes an odd number of derivatives. Here is
the only $a_2$ containing 5 derivatives, which gives rise to a consistent
cubic vertex:
\begin{eqnarray}a_2=f_{AB}C^*_\c U^A_{\a\m|\b\n|\r}
V^{B\a\m|\b\n|\c\r}d^Dx
\nonumber\end{eqnarray}
where $U^A_{\mu_1\nu_1|\mu_2\nu_2\t}=Y^2(\6^2_{\mu_1\mu_2}
C^A_{\nu_1\nu_2\t})$ and $V^A_{\mu_1\nu_1|\mu_2\nu_2|\mu_3\nu_3}
=Y^3(\6^2_{\mu_1\mu_2\mu_3}C^A_{\nu_1\nu_2\nu_3})$.\\
Then, the inhomogenous solution of $\d a_2+\g a_1=db_1$ can be computed.
The structure constants have to be symmetric in order for $a_1$ to exist :
$f_{AB}=f_{(AB)}$
\begin{eqnarray}a_1=\tilde{a}_1+\bar{a}_1=f_{(AB)}\left[
h^{*\phantom{\c}\s}_{\phantom{*}\c}\6^2_{\a\b}\phi^A_{\m\n\r\s}
V^{B\a\m|\b\n|\c\r} - 2 h^{*\g\s} \6^3_{\a\b[\c}\phi^A_{\r]\m\n\s}
U^{B\a\m|\b\n|\r} \right]d^Dx+\bar{a}_1\,.
\nonumber\end{eqnarray}
Finally, the last equation is $\d a_0+\g a_1=db_0$. It allows a solution,
unique up to redefinitions of the fields and trivial gauge transformations. We
have to say that the natural writing of $a_2$ and the vertex written in terms
of the Weyl tensor $w_{\a\b|\c\d}$ do not match automatically. In order to get
a solution, we first classified the terms of the form $w \6^4(\phi\phi)$. Then
we classified the possible terms in $\bar{a}_1$, which can be chosen in
$H^1(\g)$. So they are proportional to the field antifields, proportional to a
gauge invariant tensor ($K_{PF}$, $F_4$ or $K_4$) and proportional to a non
exact ghost. Finally, we had to introduce an arbitrary trivial combination in
order for the expressions to match. The computation cannot be made by hand
(there are thousands of terms). By using the software FORM \cite{Form}, we
managed to solve the heavy system of equations and found a consistent set of
coefficients. We obtained the following $\bar{a}_1$ as well as the vertex
$a_0$ that we wrote above (\ref{v244}):
\begin{eqnarray}\bar{a}_1=\frac{4}{D+2}\,
f_{AB}\phi^{*A\a}_{\phantom{*A\a}\b}\6^\t K^{\m\n|\a\s}
U^B_{\m\n|\b\s|\t}d^Dx-2f_{AB}\phi^{*A\m\r}_{\phantom{*A\m\r}\a\b}
\6^\t K^{\a\n|\b\s} U^B_{\m\n|\r\s|\t}d^Dx\nonumber\ .
\end{eqnarray}
Let us now consider the other possible cases for $a_2$, containing 3 or 1
derivatives.
The only possibility with three derivatives is
$a_{2,3}=g_{AB}C^*_\b \6_{[\a}C^A_{\m]\n\r}U^{B\a\m|\b\n|\r}d^Dx$.
Its variation under $\delta$ should be $\g$-closed modulo $d$ but some
nontrivial terms remain,
of the types $g_{AB}h^*U^A U^B$ and $g_{AB}h^*\6_{[.}C^A_{.]..}V^B$.
The first one can be set to zero by imposing symmetric structure constants,
but the second cannot be eliminated.
The same occurs for one of the candidates with 1 derivative:
$a_{2,1,1}=k_{AB} C^{*\b} C^{A\m\n\r}\6_{[\b} C^B_{\m]\n\r}d^Dx$.
We are then left with 2 candidates involving the spin 4 antifield.
We have found that $\d a_2+\g a_1=d b_1$ can have a solution only
if their structure constants
are proportional :
\begin{eqnarray}
a_{2,1}&=&l_{AB}C^{*A\m\n\r}\left[ C^\a
\6_{[\a} C^B_{\m]\n\r}+2\6_{[\m}C_{\a]}C^{B\a}_{\phantom{B\a}\n\r}\right]d^Dx
\nonumber\\
a_{1,1}&=&l_{AB}\phi^{*\m\n\r\s}\left[2h_\s^{\phantom{\s}\a}\6_{[\a} C^B_{\m]\n\r}
-\frac{4}{3}\,C^\a\6_{[\a}\phi^B_{\m]\n\r\s}+8\6_{[\m}h_{\a]\s}
C^{B\a}_{\phantom{B\a}\n\r}
-2\6_{[\m} C_{\a]}\phi^{B\a}_{\phantom{B\a}\n\r\s}\right]
\nonumber\\
&& +\frac{2}{D+2}l_{AB}\6_\s\phi'^{*A\r\s}C^\a\phi'^B_{\a\r}\;.\nonumber
\end{eqnarray}
There is no homogenous part $\bar{a}_1$
(because the $\g$-invariant tensors contain at least 2 derivatives).
Then, we have considered the most general expression for $a_0$, which
is a linear combination of 55 terms of the types $h\phi\6^2\phi$ and
$h\6\phi\6\phi\,$.
We have found that the equation $\d a_1+\g a_0=db_0$ does not admit any
solution.
We can conclude that the vertex found with 6 derivatives is the \emph{unique}
nonabelian $2-4-4$ cubic deformation.

\end{appendix}

\providecommand{\href}[2]{#2}\begingroup\raggedright\endgroup

\end{document}